\documentclass[trackchanges, twocolumn]{aastex701}

\newcommand{\tnsurl}{https://www.wis-tns.org/search?reporting_groupid[]=178}
\usepackage{hyperref}
\usepackage{CJKutf8}
\usepackage{soul}
\usepackage{threeparttable}
\usepackage{adjustbox}
\newcommand\N[1]{{\color{blue} \bf #1}}

\begin{document}

\title{First Results from the LSST Shadow Survey: The Restless Luminous Blue Variable AT\,2017des in the Virgo-Cluster Galaxy, NGC\,4532}

\newcommand{\LCO}{\affiliation{Las Cumbres Observatory, 6740 Cortona Drive, Suite 102, Goleta, CA 93117-5575, USA}}
\newcommand{\UCSB}{\affiliation{Department of Physics, University of California, Santa Barbara, CA 93106-9530, USA}}
\newcommand{\UCSD}{\affiliation{Department of Astronomy \& Astrophysics, University of California, San Diego, 9500 Gilman Drive, MC 0424, La Jolla, CA 92093-0424, USA}}
\newcommand{\KITP}{\affiliation{Kavli Institute for Theoretical Physics, University of California, Santa Barbara, CA 93106-4030, USA}}
\newcommand{\UCD}{\affiliation{Department of Physics and Astronomy, University of California, Davis, 1 Shields Avenue, Davis, CA 95616-5270, USA}}
\newcommand{\WIS}{\affiliation{Department of Particle Physics and Astrophysics, Weizmann Institute of Science, 76100 Rehovot, Israel}}
\newcommand{\OKC}{\affiliation{Oskar Klein Centre, Department of Astronomy, Stockholm University, Albanova University Centre, SE-106 91 Stockholm, Sweden}}
\newcommand{\OAPD}{\affiliation{INAF-Osservatorio Astronomico di Padova, Vicolo dell'Osservatorio 5, I-35122 Padova, Italy}}
\newcommand{\OAB}{\affiliation{INAF-Osservatorio Astronomico di Brera, Via E. Bianchi 46, I-23807, Merate (LC), Italy}}
\newcommand{\Caltech}{\affiliation{Cahill Center for Astronomy and Astrophysics, California Institute of Technology, Mail Code 249-17, Pasadena, CA 91125, USA}}
\newcommand{\GSFC}{\affiliation{Astrophysics Science Division, NASA Goddard Space Flight Center, Mail Code 661, Greenbelt, MD 20771, USA}}
\newcommand{\UMD}{\affiliation{Joint Space-Science Institute, University of Maryland, College Park, MD 20742, USA}}
\newcommand{\UCB}{\affiliation{Department of Astronomy, University of California, Berkeley, CA 94720-3411, USA}}
\newcommand{\TTU}{\affiliation{Department of Physics, Texas Tech University, Box 41051, Lubbock, TX 79409-1051, USA}}
\newcommand{\STScI}{\affiliation{Space Telescope Science Institute, 3700 San Martin Drive, Baltimore, MD 21218-2410, USA}}
\newcommand{\UT}{\affiliation{Department of Astronomy, The University of Texas at Austin, 2515 Speedway, Stop C1400,
Austin, TX 78712, USA}}
\newcommand{\IoA}{\affiliation{Institute of Astronomy, University of Cambridge, Madingley Road, Cambridge CB3 0HA, UK}}
\newcommand{\QUB}{\affiliation{Astrophysics Research Centre, School of Mathematics and Physics, Queen's University Belfast, Belfast BT7 1NN, UK}}
\newcommand{\IPAC}{\affiliation{IPAC, Mail Code 100-22, Caltech, 1200 E.\ California Blvd., Pasadena, CA 91125}}
\newcommand{\JPL}{\affiliation{Jet Propulsion Laboratory, California Institute of Technology, 4800 Oak Grove Dr, Pasadena, CA 91109, USA}}
\newcommand{\Southampton}{\affiliation{Department of Physics and Astronomy, University of Southampton, Southampton SO17 1BJ, UK}}
\newcommand{\LANL}{\affiliation{Space and Remote Sensing, MS B244, Los Alamos National Laboratory, Los Alamos, NM 87545, USA}}
\newcommand{\Tsinghua}{\affiliation{Physics Department and Tsinghua Center for Astrophysics, Tsinghua University, Beijing, 100084, People's Republic of China}}
\newcommand{\NAOC}{\affiliation{National Astronomical Observatory of China, Chinese Academy of Sciences, Beijing, 100012, People's Republic of China}}
\newcommand{\Itagaki}{\affiliation{Itagaki Astronomical Observatory, Yamagata 990-2492, Japan}}
\newcommand{\Einstein}{\altaffiliation{Einstein Fellow}}
\newcommand{\Hubble}{\altaffiliation{Hubble Fellow}}
\newcommand{\CfA}{\affiliation{Center for Astrophysics \textbar{} Harvard \& Smithsonian, 60 Garden Street, Cambridge, MA 02138-1516, USA}}
\newcommand{\UA}{\affiliation{Steward Observatory, University of Arizona, 933 North Cherry Avenue, Tucson, AZ 85721-0065, USA}}
\newcommand{\MPIA}{\affiliation{Max-Planck-Institut f\"ur Astrophysik, Karl-Schwarzschild-Stra\ss{}e 1, D-85748 Garching, Germany}}
\newcommand{\DSFP}{\altaffiliation{LSSTC Data Science Fellow}}
\newcommand{\HCO}{\affiliation{Harvard College Observatory, 60 Garden Street, Cambridge, MA 02138-1516, USA}}
\newcommand{\Carnegie}{\affiliation{Observatories of the Carnegie Institute for Science, 813 Santa Barbara Street, Pasadena, CA 91101-1232, USA}}
\newcommand{\TAU}{\affiliation{School of Physics and Astronomy, Tel Aviv University, Tel Aviv 69978, Israel}}
\newcommand{\Edinburgh}{\affiliation{Institute for Astronomy, University of Edinburgh, Royal Observatory, Blackford Hill EH9 3HJ, UK}}
\newcommand{\Birmingham}{\affiliation{Birmingham Institute for Gravitational Wave Astronomy and School of Physics and Astronomy, University of Birmingham, Birmingham B15 2TT, UK}}
\newcommand{\Bath}{\affiliation{Department of Physics, University of Bath, Claverton Down, Bath BA2 7AY, UK}}
\newcommand{\CTIO}{\affiliation{Cerro Tololo Inter-American Observatory, National Optical Astronomy Observatory, Casilla 603, La Serena, Chile}}
\newcommand{\Potsdam}{\affiliation{Leibniz-Institut f\"ur Astrophysik Potsdam (AIP), An der Sternwarte 16, D-14482 Potsdam, Germany}}
\newcommand{\INPE}{\affiliation{Instituto Nacional de Pesquisas Espaciais, Avenida dos Astronautas 1758, 12227-010, S\~ao Jos\'e dos Campos -- SP, Brazil}}
\newcommand{\UNC}{\affiliation{Department of Physics and Astronomy, University of North Carolina, 120 East Cameron Avenue, Chapel Hill, NC 27599, USA}}
\newcommand{\Ohio}{\affiliation{Astrophysical Institute, Department of Physics and Astronomy, 251B Clippinger Lab, Ohio University, Athens, OH 45701-2942, USA}}
\newcommand{\AAS}{\affiliation{American Astronomical Society, 1667 K~Street NW, Suite 800, Washington, DC 20006-1681, USA}}
\newcommand{\MMT}{\affiliation{MMT and Steward Observatories, University of Arizona, 933 North Cherry Avenue, Tucson, AZ 85721-0065, USA}}
\newcommand{\Geneva}{\affiliation{ISDC, Department of Astronomy, University of Geneva, Chemin d'\'Ecogia, 16 CH-1290 Versoix, Switzerland}}
\newcommand{\IUCAA}{\affiliation{Inter-University Center for Astronomy and Astrophysics, Post Bag 4, Ganeshkhind, Pune, Maharashtra 411007, India}}
\newcommand{\CMU}{\affiliation{Department of Physics, Carnegie Mellon University, 5000 Forbes Avenue, Pittsburgh, PA 15213-3815, USA}}
\newcommand{\NAOJ}{\affiliation{Division of Science, National Astronomical Observatory of Japan, 2-21-1 Osawa, Mitaka, Tokyo 181-8588, Japan}}
\newcommand{\IfA}{\affiliation{Institute for Astronomy, University of Hawai`i, 2680 Woodlawn Drive, Honolulu, HI 96822-1839, USA}}
\newcommand{\UCSC}{\affiliation{Department of Astronomy and Astrophysics, University of California, Santa Cruz, CA 95064-1077, USA}}
\newcommand{\Purdue}{\affiliation{Department of Physics and Astronomy, Purdue University, 525 Northwestern Avenue, West Lafayette, IN 47907-2036, USA}}
\newcommand{\Princeton}{\affiliation{Department of Astrophysical Sciences, Princeton University, 4 Ivy Lane, Princeton, NJ 08540-7219, USA}}
\newcommand{\Moore}{\affiliation{Gordon and Betty Moore Foundation, 1661 Page Mill Road, Palo Alto, CA 94304-1209, USA}}
\newcommand{\Durham}{\affiliation{Department of Physics, Durham University, South Road, Durham, DH1 3LE, UK}}
\newcommand{\JHU}{\affiliation{Department of Physics and Astronomy, The Johns Hopkins University, 3400 North Charles Street, Baltimore, MD 21218, USA}}
\newcommand{\Toronto}{\affiliation{David A.\ Dunlap Department of Astronomy and Astrophysics, University of Toronto,\\ 50 St.\ George Street, Toronto, Ontario, M5S 3H4 Canada}}
\newcommand{\Duke}{\affiliation{Department of Physics, Duke University, Campus Box 90305, Durham, NC 27708, USA}}
\newcommand{\NCU}{\affiliation{Graduate Institute of Astronomy, National Central University, 300 Jhongda Road, 32001 Jhongli, Taiwan}}
\newcommand{\Columbia}{\affiliation{Department of Physics and Columbia Astrophysics Laboratory, Columbia University, Pupin Hall, New York, NY 10027, USA}}
\newcommand{\Flatiron}{\affiliation{Center for Computational Astrophysics, Flatiron Institute, 162 5th Avenue, New York, NY 10010-5902, USA}}
\newcommand{\NU}{\affiliation{Department of Physics and Astronomy, Northwestern University, 2145 Sheridan Rd, Evanston, IL 60208, USA}}
\newcommand{\CIERA}{\affiliation{Center for Interdisciplinary Exploration and Research in Astrophysics, \\Northwestern University, 1800 Sherman Avenue, 8th Floor, Evanston, IL 60201, USA}}
\newcommand{\SkAI}{\affil{NSF-Simons AI Institute for the Sky (SkAI), 172 E. Chestnut St., Chicago, IL 60611, USA}}
\newcommand{\GeminiNorth}{\affiliation{Gemini Observatory, 670 North A`ohoku Place, Hilo, HI 96720-2700, USA}}
\newcommand{\Keck}{\affiliation{W.~M.~Keck Observatory, 65-1120 M\=amalahoa Highway, Kamuela, HI 96743-8431, USA}}
\newcommand{\UW}{\affiliation{Department of Astronomy, University of Washington, 3910 15th Avenue NE, Seattle, WA 98195-0002, USA}}
\newcommand{\Catalyst}{\altaffiliation{LSST-DA Catalyst Fellow}}
\newcommand{\USask}{\affiliation{Department of Physics and Engineering Physics, University of Saskatchewan, 116 Science Place, Saskatoon, SK S7N 5E2, Canada}}
\newcommand{\Thacher}{\affiliation{Thacher School, 5025 Thacher Road, Ojai, CA 93023-8304, USA}}
\newcommand{\Rutgers}{\affiliation{Department of Physics and Astronomy, Rutgers, the State University of New Jersey,\\136 Frelinghuysen Road, Piscataway, NJ 08854-8019, USA}}
\newcommand{\FSU}{\affiliation{Department of Physics, Florida State University, 77 Chieftan Way, Tallahassee, FL 32306-4350, USA}}
\newcommand{\Melbourne}{\affiliation{School of Physics, The University of Melbourne, Parkville, VIC 3010, Australia}}
\newcommand{\ASTROthreeD}{\affiliation{ARC Centre of Excellence for All Sky Astrophysics in 3 Dimensions (ASTRO 3D)}}
\newcommand{\Stromlo}{\affiliation{Mt.\ Stromlo Observatory, The Research School of Astronomy and Astrophysics, Australian National University, ACT 2601, Australia}}
\newcommand{\NCPAS}{\affiliation{National Centre for the Public Awareness of Science, Australian National University, Canberra, ACT 2611, Australia}}
\newcommand{\TAMU}{\affiliation{Department of Physics and Astronomy, Texas A\&M University, 4242 TAMU, College Station, TX 77843, USA}}
\newcommand{\Mitchell}{\affiliation{George P.\ and Cynthia Woods Mitchell Institute for Fundamental Physics \& Astronomy, College Station, TX 77843, USA}}
\newcommand{\ESO}{\affiliation{European Southern Observatory, Alonso de C\'ordova 3107, Casilla 19, Santiago, Chile}}
\newcommand{\ICE}{\affiliation{Institute of Space Sciences (ICE, CSIC), Campus UAB, Carrer
de Can Magrans, s/n, E-08193 Barcelona, Spain}}
\newcommand{\IEEC}{\affiliation{Institut d'Estudis Espacials de Catalunya, Gran Capit\`a, 2-4, Edifici Nexus, Desp.\ 201, E-08034 Barcelona, Spain}}
\newcommand{\Warwick}{\affiliation{Department of Physics, University of Warwick, Gibbet Hill Road, Coventry CV4 7AL, UK}}
\newcommand{\Macquarie}{\affiliation{School of Mathematical and Physical Sciences, Macquarie University, NSW 2109, Australia}}
\newcommand{\AAARC}{\affiliation{Astronomy, Astrophysics and Astrophotonics Research Centre, Macquarie University, Sydney, NSW 2109, Australia}}
\newcommand{\Capodimonte}{\affiliation{INAF - Capodimonte Astronomical Observatory, Salita Moiariello 16, I-80131 Napoli, Italy}}
\newcommand{\INFNNapoli}{\affiliation{INFN - Napoli, Strada Comunale Cinthia, I-80126 Napoli, Italy}}
\newcommand{\ICRANet}{\affiliation{ICRANet, Piazza della Repubblica 10, I-65122 Pescara, Italy}}
\newcommand{\MSU}{\affiliation{Center for Data Intensive and Time Domain Astronomy, Department of Physics and Astronomy,\\Michigan State University, East Lansing, MI 48824, USA}}
\newcommand{\IAP}{\affiliation{Institut d'Astrophysique de Paris, CNRS-Sorbonne Universit\'e, 98 bis boulevard Arago, 75014 Paris, France}}
\newcommand{\Pitt}{\affiliation{Department of Physics and Astronomy \& Pittsburgh Particle Physics, Astrophysics, and Cosmology Center (PITT PACC), University of Pittsburgh, 3941 O'Hara Street, Pittsburgh, PA 15260, USA}}
\newcommand{\Vtech}{\affiliation{Department of Physics, Virginia Tech, Blacksburg, VA 24061, USA}}
\newcommand{\IAC}{\affiliation{Instituto de Astrof{\'\i}sica de Canarias, E-38205 La Laguna, Tenerife, Spain}}
\newcommand{\Laguna}{\affiliation{Universidad de La Laguna, Dept. Astrof{\'\i}sica, E-38206 La Laguna, Tenerife, Spain}}
\newcommand{\UOak}{\affiliation{Homer L. Dodge Department of Physics and Astronomy, University of Oklahoma, 440 W. Brooks, Rm 100, Norman, OK 73019-2061, USA}}
\newcommand{\Hamburg}{\affiliation{Hamburger Sternwarte, Gojenbergsweg 112, D-21029 Hamburg, Germany}}
\newcommand{\PSI}{\affiliation{Planetary Science Institute, 1700 East Fort Lowell, Suite 106, Tucson, AZ 85719-2395 USA}}
\newcommand{\SETI}{\affiliation{SETI Institute, 339 Bernardo Ave, Suite 200, Mountain View, CA 94043, USA}}
\newcommand{\Hobart}{\affiliation{Physics Department, Hobart and William Smith Colleges, 300 Pulteney Street, Geneva, NY 14456, USA}}
\newcommand{\Cornell}{\affiliation{Department of Astronomy, Cornell University, 245 East Avenue, Ithaca, NY 14850, USA}}
\newcommand{\Athens}{\affiliation{IAASARS, National Observatory of Athens, Penteli 15236, Greece}}
\newcommand{\Turki}{\affiliation{Department of Physics and Astronomy, University of Turku, Vesilinnantie 5, 20500 Finland}}
\newcommand{\UMNAstro}{\affiliation{School of Physics and Astronomy, University of Minnesota, 116 Church Street S.E., Minneapolis, MN 55455, USA}}
\newcommand{\UVa}{\affiliation{Department of Astronomy, 530 McCormick Road, Charlottesville, VA 22904-4325, USA}}
\newcommand{\UTA}{\affiliation{Department of Physics, University of Texas at Arlington, Box 19059, Arlington, TX 76019, USA}}
\newcommand{\Konkoly}{\affiliation{Konkoly Observatory, CSFK, MTA Center of Excellence, Konkoly-Thege M. \'ut 15-17, Budapest, 1121, Hungary}}
\newcommand{\ELTE}{\affiliation{ELTE E\"otv\"os Lor\'and University, Institute of Physics and Astronomy, P\'azm\'any P\'eter s\'et\'any 1/A, Budapest, 1117 Hungary}}
\newcommand{\Szeged}{\affiliation{Department of Experimental Physics, University of Szeged, D\'om t\'er 9, Szeged, 6720, Hungary}}
\newcommand{\IAIFI}{\affiliation{The NSF AI Institute for Artificial Intelligence and Fundamental Interactions, USA}}
\newcommand{\UPadua}{\affiliation{Physics and Astronomy Department Galileo Galilei, University of Padova, Vicolo dell'Osservatorio 3, I-35122, Padova, Italy}}
\newcommand{\UWarwick}{\affiliation{Department of Physics, Gibbet Hill Road, University of Warwick, Coventry CV4 7AL, United Kingdom}}
\newcommand{\ING}{\affiliation{Isaac Newton Group of Telescopes, Apt. de Correos 368, E-38700 Santa Cruz de la Palma, Spain}}
\newcommand{\UNott}{\affiliation{School of Physics and Astronomy, University of Nottingham, University Park, Nottingham, NG7 2RD}}
\newcommand{\USurrey}{\affiliation{Mullard Space Science Laboratory, University College London, Holmbury St Mary, Dorking, Surrey RH5 6NT, United Kingdom}}
\newcommand{\USheffiled}{\affiliation{Department of Physics and Astronomy, University of Sheffield, Sheffield S3 7RH, UK}}
\newcommand{\Adler}{\affiliation{Adler Planetarium, 1300 S. DuSable Lake Shore Dr., Chicago, IL 60605, USA}}
\newcommand{\Monash}{\affiliation{School of Physics and Astronomy, Monash University, Clayton, Victoria 3800, Australia}}
\newcommand{\OzGrav}{\affiliation{OzGrav: The ARC Centre of Excellence for Gravitational Wave Discovery, Clayton, Victoria 3800, Australia}}
\newcommand{\Christ}{\affiliation{School of Physical and Chemical Sciences -- Te Kura Mat\={u}, University of Canterbury, Private Bag 4800, Christchurch 8140, \\ Aotearoa, New Zealand}}
\newcommand{\mta}{\affiliation{MTA-ELTE Lendület “Momentum” Milky Way Research Group, Szent Imre H. st. 112, 9700 Szombathely, Hungary}}
\newcommand{\bao}{\affiliation{Baja Astronomical Observatory of University of Szeged, Szegedi {\'u}t,
Kt. 766, 6500 Baja, Hungary}}
\newcommand{\hrs}{\affiliation{HUN-REN--SZTE Stellar Astrophysics Research Group, Szegedi {\'u}t, Kt.
766, 6500 Baja, Hungary}}
\newcommand{\noir}{\affiliation{NSF NOIRLab, 950 N Cherry Ave, Tucson, AZ 85719, USA}}
\newcommand{\noirctio}{\affiliation{Cerro Tololo Inter-American Observatory/NSF’s NOIRLab, Casilla 603, La Serena, Chile}}
\newcommand{\UVA}{\affiliation{Department of Astronomy, University of Virginia, Charlottesville, VA22903, USA}}
\newcommand{\UF}{\affiliation{Department of Astronomy, University of Florida, Bryant Space Science Center, Gainesville, FL
32611-2055, USA}}
\newcommand{\noirHI}{\affiliation{NSF NOIRLab, 670 N. A’ohoku Place, Hilo, Hawai’i, 96720, USA}}
\newcommand{\hunren}{\affiliation{HUN-REN CSFK, Konkoly Observatory, MTA Centre of Excellence, Konkoly Thege Miklós út 15-17, Budapest, 1121, Hungary}}
\newcommand{\ljmu}{\affiliation{Astrophysics Research Institute, Liverpool John Moores University, IC2, 146 Brownlow Hill, Liverpool, L3 5RF, UK}}

% \correspondingauthor{Conor~L.~Ransome}
% \email{cransome@arizona.edu}
\author[orcid=0000-0003-4175-4960]{Conor~L.~Ransome}
\UA
%\affiliation{University of Saskatchewan}
\email[show]{cransome@arizona.edu} 

\author[0000-0001-8073-8731]{Bhagya M.\ Subrayan}
\UA
\email{bsubrayan@arizona.edu}

\author[0000-0003-4102-380X]{David J.\ Sand}
\UA
\email{dsand@arizona.edu}

\author[0000-0002-9454-1742]{Brian Hsu}
\UA
\email{bhsu@arizona.edu}

\author[0000-0002-9364-5419]{Xander J. Hall}
\CMU
\email{xjh@andrew.cmu.edu}

\author[0000-0002-0744-0047]{Jeniveve Pearson}
\UA
\email{jenivevepearson@arizona.edu}

\author[0000-0002-4924-444X]{K.\ Azalee Bostroem}
\Catalyst\UA
\email{bostroem@arizona.edu}

\author[0000-0003-0123-0062]{Jennifer E.\ Andrews}
\GeminiNorth
\email{jennifer.andrews@noirlab.edu}

\author[0000-0001-8764-7832]{Jozsef Vinko}
\hunren\Szeged
\UT
\email{vinko.jozsef.@csfk.org}

\author[0000-0003-1349-6538]{J. Craig Wheeler}
\UT
\email{wheel@astro.as.utexas.edu}

\author[0009-0006-0647-636X]{Phillip Noel}
\UA
\email{phillipnoel@arizona.edu}

\author[0000-0001-7201-1938]{Lei Hu}
\CMU
\email{leihu@andrew.cmu.edu}

\author[0000-0002-1270-7666]{Tom\'as Cabrera}
\CMU
\email{tcabrera@andrew.cmu.edu}

\author[0000-0001-8818-0795]{Stefano Valenti}
\UCD
\email{stfn.valenti@gmail.com}

\author[0000-0002-3934-2644]{Wynn V. Jacobson-Gal\'an}
\altaffiliation{NASA Hubble Fellow}
\Caltech
\email{wynnjg@caltech.edu}

\author[0000-0001-5510-2424]{Nathan Smith}
\UA
\email{nathans@as.arizona.edu}

\author[0000-0003-3460-0103]{Alexei V. Filippenko}
\UCB
\affiliation{Hagler Institute for Advanced Study, Texas A\&M University, 3572 TAMU, College Station, TX 77843, USA}
\email{afilippenko@berkeley.edu}

%% ALPHABETICAL %%

\author[0000-0001-8341-3940]{Mojgan Aghakhanloo}
\affiliation{Department of Astronomy, University of Virginia, 530 McCormick Road, Charlottesville, VA 22904, USA}
\affiliation{Virginia Institute for Theoretical Astronomy, University of Virginia, Charlottesville, VA 22904, USA}
\email{}

\author[0000-0002-8977-1498]{Igor Andreoni}
\UNC
\email{igor.andreoni@unc.edu}

\author[0000-0002-1895-6639]{Moira Andrews}
\LCO\UCSB
\email{mandrews@lco.global}

\author[0000-0001-7090-4898]{Iair Arcavi}
\TAU
\email{arcavi@tauex.tau.ac.il}

\author[0000-0002-4449-9152]{Katie~Auchettl}
\affiliation{Department of Astronomy and Astrophysics, University of California, Santa Cruz, CA 95064, USA}
\affiliation{School of Physics, The University of Melbourne, VIC 3010, Australia}
\email{}

\author[0009-0004-7268-7283]{Raphael Baer-Way}
\affiliation{Department of Astronomy, University of Virginia,
 Charlottesville VA 22904-4325, USA}
\email{placeholder@gmail.com}

\author[0000-0003-4666-4606]{Emma R. Beasor}
\ljmu
\email{e.r.beasor@ljmu.ac.uk}

\author[0000-0002-9392-9681]{Edo Berger}
\CfA
\email{eberger@cfa.harvard.edu}

\author[orcid=0000-0003-1953-8727]{Federica B. Bianco}
\affiliation{Department of Physics and Astronomy, University of Delaware, Newark, DE 19716, USA}
\affiliation{Joseph R. Biden, Jr. School of Public Policy and Administration, University of Delaware, DE 19716, USA}
\affiliation{University of Delaware, Data Science Institute, Newark, DE 19716, USA}
\affiliation{Vera C. Rubin Observatory, Tucson, AZ 85719, USA}
\email{}

\author[0000-0003-0526-2248]{Peter Blanchard}
\CfA
\email{pblanchard@cfa.harvard.edu}

\author[0000-0001-5955-2502]{Thomas G. Brink}
\UCB
\email{tgbrink@berkeley.edu}

\author[orcid=0000-0001-5078-5457,gname=Siddharth, sname=Chaini]{Siddharth Chaini} 
\altaffiliation{NASA FINESST Fellow}
\affiliation{Department of Physics and Astronomy, University of Delaware, Newark, DE 19716, USA}
\affiliation{University of Delaware, Data Science Institute, Newark, DE 19716, USA}
\email{chaini@udel.edu}

\author[0000-0002-7627-4839]{Adrian Crawford}
\UVA
\email{adrian.crawford@virginia.edu}

\author[0000-0002-7937-6371]{Yize Dong \begin{CJK}{UTF8}{gbsn}(董一泽)\end{CJK}}
\CfA
\email{yize.dong@cfa.harvard.edu}

\author[0000-0003-4914-5625]{Joseph Farah}
\LCO\UCSB
\email{jfarah@lco.global}

\author[0000-0003-4537-3575]{Noah Franz}
\UA
\email{nfranz@arizona.edu}

\author[0000-0001-6395-6702]{Sebastian Gomez}
\UT
\email{sebastian.gomez@austin.utexas.edu}

\author[0000-0002-9154-3136]{Melissa L.\ Graham}
\affiliation{University of Washington, Dept. of Astronomy, Box 351580, Seattle, WA 98195, USA}
\affiliation{Institute for Data-intensive Research in Astrophysics and Cosmology, University of Washington, 3910 15th Avenue NE, Seattle, WA 98195, USA}
\email{mlg3k@uw.edu}

\author[0000-0002-1125-9187]{Daichi Hiramatsu}
\UF
\email{dhiramatsu@ufl.edu}

\author[0009-0008-2052-8474]{\'Agoston Horti-D\'avid}
\affiliation{ELTE E\"otv\"os Lor\'and University, Institute of Physics and Astronomy, 
P\'azm\'any P\'eter s\'et\'any 1A, Budapest 1117, Hungary }
\hunren
\email{hortidavid.agoston@csfk.org}

\author[0000-0002-0832-2974]{Griffin Hosseinzadeh}
\UCSD
\email{ghosseinzadeh@ucsd.edu}

\author[0000-0003-4253-656X]{D.\ Andrew Howell}
\LCO\UCSB
\email{ahowell@lco.global}

\author[0000-0003-4131-5183]{Philip A. James}
\ljmu
\email{P.A.James@ljmu.ac.uk}

\author[0000-0001-8738-6011]{Saurabh W.\ Jha}
\Rutgers
\email{saurabh@physics.rutgers.edu}

\author[0000-0002-5740-7747]{Charles D. Kilpatrick}
\CIERA
\email{ckilpatrick@northwestern.edu}

\author[0000-0003-3108-1328]{Lindsey~A.~Kwok}
\CIERA
\email{lindsey.kwok@northwestern.edu}

\author[0000-0001-5169-4143]{Gavin P. Lamb}
\ljmu
\email{G.P.Lamb@ljmu.ac.uk}

\author[0000-0002-6164-5051]{Amanda~R.~Lopes}
\affiliation{Instituto de Astronomia, Geofísica e Ciências Atmosféricas da Universidade de São Paulo, Cidade Universitária, CEP:05508-990, São Paulo, SP, Brazil}
\email[]{}

%Acknowledgments: A.R.L. acknowledges the support grant 2025/09544-0 from São Paulo Research Foundation (FAPESP).

\author[0000-0001-9589-3793]{Michael Lundquist}
\Keck
\email{mlundquist@keck.hawaii.edu}

\author[0000-0002-9144-7726]{Clara E. Martínez-Vázquez}
\noirHI
\email{ckilpatrick@northwestern.edu}

\author[0000-0001-6685-0479]{Thomas Matheson}
\noir
\email{tom.matheson@noirlab.edu}

\author[0000-0001-5807-7893]{Curtis McCully}
\LCO
\email{cmccully@lco.global}

\author[0000-0001-7132-0333]{Maryam Modjaz}
\affiliation{Department of Astronomy, University of Virginia, 530 McCormick Road, Charlottesville, VA 22904, USA}
\email{mmodjaz@virginia.edu}

\author[orcid=0000-0002-6639-6533,gname='Gregory', sname='Paek']{Gregory S. H. Paek}  
\affiliation{Institute for Astronomy, University of Hawai`i, 2680 Woodlawn Drive, Honolulu, HI 96822, USA}
\email{gregorypaek94@gmail.com}

\author[0000-0002-6011-0530]{Antonella Palmese}
\CMU
\email{tcabrera@andrew.cmu.edu}

\author[0009-0000-5120-1193]{Avi Patel}
\UCSC
\email{avpapate@ucsc.edu}

\author[0000-0001-8907-3051]{Joanne L.\ Pledger}
\affiliation{Jeremiah Horrocks Institute, University of Lancashire, Preston, PR1 2HE, UK}
\email{jpledger@lancashire.ac.uk}

\author[0000-0002-7352-7845]{Aravind P.\ Ravi}
\UCD
\email{apazhayathravi@ucdavis.edu}

\author[0000-0003-3643-839X]{Jeonghee Rho}
\SETI
\email{jrho@seti.org}

\author[0000-0003-0926-3950]{Kriszti\'an S\'arneczky}
\hunren
\email{sarneczky.krisztian@csfk.org}

\author[0000-0002-4022-1874]{Manisha Shrestha}
\Monash\OzGrav
\email{Manisha.Shrestha@monash.edu}

\author[0000-0002-3346-2370]{Richard Smith}
\affiliation{Department of Astronomy \& Astrophysics, University of Chicago, 5640 South Ellis Avenue, Chicago, IL 60637, USA}
\email{rsmith25@uchicago.edu}

\author[]{Analía V. Smith Castelli}
\affiliation{Instituto de Astrofísica de La Plata (CONICET - UNLP), Paseo del Bosque s/n, B1900FWA, La Plata, Argentina, Facultad de Ciencias Astronómicas y Geofísicas, Universidad Nacional de La Plata, Paseo del Bosque s/n, B1900FWA, La Plata, Argentina}
\email[]{}

\author[0000-0001-6360-992X]{Monika Soraisam}
\noirHI
\email{monika.soraisam@noirlab.edu}

\author[0000-0002-1468-9668]{Jay Strader}
\MSU
\email{straderj@msu.edu}

\author[0000-0001-5567-1301]{Francisco Valdes}
\noir
\email{frank.valdes@noirlab.edu}

\author[0000-0002-4951-8762]{Sergiy Vasylyev}
\UCSD
\email{svasylyev@ucsd.edu}

\author[0000-0002-5814-4061]{V.~Ashley~Villar}
\affiliation{Center for Astrophysics \textbar{} Harvard \& Smithsonian, 60 Garden Street, Cambridge, MA 02138-1516, USA}
\affiliation{The NSF AI Institute for Artificial Intelligence and Fundamental Interactions}
\email{ashleyvillar@cfa.harvard.edu}

\author[0000-0003-4341-6172]{A.~Katherina~Vivas}
\noirctio
\email{kathy.vivas@noirlab.edu}

\author[0000-0001-7092-9374]{Lifan Wang}
\TAMU
\email{lifan@tamu.edu}

\author[0000-0003-4537-3575]{Samuel D. Wyatt}
\GSFC
\email{samuel.d.wyatt@nasa.gov}

\author[orcid=0009-0006-7296-728X]{Kathryn Wynn}
\LCO \UCSB 
\email{}

\author[0000-0001-5955-2502]{WeiKang Zheng}
\UCB
\email{weikang@berkeley.edu}

% Department of Experimental Physics, Institute of Physics, University of
% Szeged, D{\'o}m t{\'e}r 9, 6720 Szeged, Hungary
% \and
% Baja Astronomical Observatory of University of Szeged, Szegedi {\'u}t,
% Kt. 766, 6500 Baja, Hungary
% \and
% HUN-REN--SZTE Stellar Astrophysics Research Group, Szegedi {\'u}t, Kt.
% 766, 6500 Baja, Hungary

% \author[0009-0003-8380-4003]{Zachary G. Lane}
% \Christ
% \email{zachary.lane@pg.canterbury.ac.nz}

% \author[0009-0008-9693-4348]{Darshana Mehta}
% \UCD
% \email{ddmehta@ucdavis.edu}
%\author[0000-0002-0370-157X]{Peter Milne}
%\UA

% \author[0000-0001-9570-0584]{Megan Newsome}
% \UT
% \email{newsome.megane@gmail.com}

% \author[0000-0003-0209-9246]{Estefania Padilla-Gonzalez}
% \JHU
% \email{epadill7@jh.edu}

% \author[0000-0002-7015-3446]{Nicolas E.\ Meza Retamal}
% \UCD
% \email{nemezare@ucdavis.edu}

% \author[0000-0003-4610-1117]{Tam\'as Szalai}
% \Szeged\mta
% \email{szaszi@titan.physx.u-szeged.hu}

% \author[0000-0002-4951-8762]{Sergiy Vasylyev}
% \UCSD
% \email{svasylyev@ucsd.edu}

% \author[0009-0001-3106-0917]{Aidan Martas}
% \UCB \UCD
% \email{aidmart@berkeley.edu}

% \author[0000-0002-7866-4531]{Chang~Liu}
% \NU
% \CIERA
% \SkAI
% \email{ptg.cliu@u.northwestern.edu}

% \author[0000-0003-0549-3281]{Daryl Janzen}
% \USask
% \email{daryl.janzen@usask.ca}

% \author[0000-0001-5754-4007]{Jacob E.\ Jencson}
% \IPAC
% \email{jjencson@ipac.caltech.edu}

%\author[0000-0003-3460-0103]{Alexei V. Filippenko}
\UCB
% \email{alex@astro.berkeley.edu}

% \author[0000-0003-4341-9824]{Kiranjyot Gill}
% \affiliation{Independent Researcher}
% \email{kiranjyotkgill@gmail.com}

% \author[0000-0003-2744-4755]{Emily T. Hoang}
% \UCD
% \email{emthoang@ucdavis.edu}

% \author[0000-0003-0528-202X]{Collin Christy}
% \UA
% \email{collinchristy@arizona.edu}

% \author{Istv{\'a}n Cs{\'a}nyi}
% \bao 
% \email{}

% \author[0000-0002-1481-4676]{Samaporn~Tinyanont}
% \affiliation{National Astronomical Research Institute of Thailand, 260 Moo 4, Donkaew, Maerim, Chiang Mai, 50180, Thailand}
% \email{samaporn@narit.or.th}vinko.jozsef.@csfk.org

%\author[0000-0002-8297-2473]{Kate D. Alexander}
%\UA

% \author[0009-0000-9929-7518]{Dominik B{\'a}nhidi} 
% \Szeged\bao
% \email{}

% \author{Imre Barna B{\'i}r{\'o}} 
% \bao\hrs
% \email{}

% \author[0000-0002-2028-9329]{Anya Nugent}
% \CfA
% \email{anya.nugent@cfa.harvard.edu}

\collaboration{all}{The Shadow Collaboration}

\begin{abstract}

The Legacy Survey of Space and Time (LSST) will start in late-summer 2026, revolutionizing transient astronomy. Here, we present the Dark Energy Camera (DECam) Shadow Survey, which is designed to maximize the science potential of LSST by shadowing LSST observations of local galaxy-cluster fields, producing a nightly cadence of these fields. The Shadow Survey will discover extremely young supernovae (SNe), SN precursors, as well as other explosive transients and exotic phenomena, helping to characterize such transients at unprecedented cadence and depth when combined with LSST. We describe our workflow, pipeline, public data releases, and candidate vetting. As an early result of Shadow, we present the fitful luminous blue variable (LBV) eruptions of AT\,2017des in the Virgo-Cluster galaxy NGC\,4532. AT\,2017des has short-timescale variability (of order $10$ days), peaking at around $M_r=-12.5$\,mag, brighter than normal LBVs, and similar to the more extreme flaring of hot LBVs/SN impostors such as SN\,2000ch, AT\,2016blu, and the precursor activity of SN\,2009ip. Our spectral time-series reveals features typical of these hot LBVs and SN impostors/precursors. Combining our data with long-baseline photometry from additional observatories, we find that the peaks of the outbursts of AT\,2017des are getting brighter over time, with 2026 peak fluxes being up to 5 times greater than in 2023 and an average brightening of $\sim0.05$\,mag\,yr$^{-1}$. The peaks of AT\,2017des are more luminous than those of most other LBVs, only being fainter than bright precursors such as SN\,2009ip, and extreme SN impostors such as AT\,2016blu. AT\,2017des may therefore be ``ramping up'' to a terminal explosion.

%which is designed to maximize the discovery power and science potential of LSST. Our survey uses DECam to shadow LSST observations of local galaxy-cluster fields, producing nightly observations of these target-rich pointings before LSST has started in earnest, the Zwicky Transient Facility and data from Las Cumbres Observatory and Konkoly Observatory
\end{abstract}

\keywords{Circumstellar matter (241), Luminous blue variable stars (944), Surveys (1671), Variable stars (1760)}


\section{Introduction} \label{sec:intro}

Transient astronomy is on the cusp of a revolution with the dawn of the Legacy Survey of Space and Time \citep[LSST;][]{LSST2} which will commence in late-summer 2026. LSST builds on previous surveys that have produced over 10$^4$ spectroscopically classified supernovae (SNe; many more are unclassified). From \citet{Zwicky_1965} to the Zwicky Transient Facility \citep[ZTF;][]{ztf}, systematic transient surveys such as the Cal{\'a}n/Tololo Supernova Survey \citep[][]{CTS}, the Lick Observatory Supernova Survey \citep[][]{LOSS}, the Palomar Transient Factory \citep[][]{Law_2009}, the All-Sky Automated Survey for Supernovae \citep[][]{Kochanek_2017}, the Young Supernova Experiment \citep[][]{YSE}, the Gravitational-Wave Optical Transient Observer \citep[GOTO;][]{Steeghs_2022, Dyer_2024}, and the Distance Less Than 40\,Mpc Survey \citep[][]{Tartaglia_2017} have shaped the transient discovery landscape. Current and recent-past surveys, such as ZTF, are all-sky, while others observe a set of fields. LSST, in contrast, will observe the entire southern sky roughly every three days \citep{Bianco_2022}.

%From the superluminous ASASSN-15lh \citep[][]{Dong_2016} to the more modest classical novae, t

These surveys have discovered numerous SNe of all classes, as well as other transient types. The large sample sizes enabled by transient surveys have illuminated a zoo of exploding transients \citep[see, e.g.,][]{Smith_2017_review,Jha_2017, Maguire_2017, Arcavi_2017, Howell_2017,Jha_2019, Modjaz_2019, Inserra_2019, Darnley_2019_review, Cai_2021}. Apart from the usual suspects, (SNe and classical novae), this transient menagerie includes exotic, rare phenomena we only have a chance of observing using large surveys. These rare transients include intermediate-luminosity red transients \citep[e.g.,][]{Cai_2021}, fast blue optical transients \citep[][]{Perley_2019_cow, Ho_2023}, and calcium-rich transients \citep[][]{Filippenko_2003_CaR,Kasliwal_2012, Yadavalli_2024, Ravi_2026}.

Lurking between classical novae and SNe in the explosive transient timescale-luminosity phase space \citep[][]{kasphd, Villar_2017} is another zoo of objects known collectively as ``gap transients.'' Among them  are the outbursts of luminous blue variables (LBVs; see \citealt{smith26} for a review). LBVs are massive, evolved stars that exhibit variability in the form of decades-long S\,Doradus-like pulsations (with peaks at $M_V \approx -10$ to $-12$\,mag), and more violent ``giant eruption'' episodes \citep{Conti_1984,Humphreys_1994, vanGenderen_2001, Clark_2005, Smith_2011, smith26}. Galactic and Magellanic Cloud examples of LBVs include the aforementioned S\,Doradus, P\,Cygni, and $\eta$ Carinae. $\eta$\,Carinae exemplifies the extreme outbursts observed from LBVs with the 19th century Great Eruption which peaked at $M_V \approx-14$\,mag \citep{Smith_2011_etacar,smith18}, compared to the quiescent LBV luminosity of $M_V \approx-8$ to $-10\,$mag \citep{vanGenderen_2001}.

While normal S\,Doradus outbursts are not important mass-loss episodes \citep{smith26}, giant eruptions like that of $\eta$\,Carinae are observed to shed enormous amounts of material from the envelope of the star. Proposed mechanisms range from continuum-driven super-Eddington winds \citep{Owocki_2004} to violent binary interactions \citep{Soker_2006,Kashi_2010,Smith_2011_binlbv,Kashi_2013,Soker_2013} or stellar merger events \citep{smith18}. The mass-loss rates of LBVs range from 10$^{-5}$\,M$_\odot$\,yr$^{-1}$ for LBV winds \citep{Leitherer_1997}, to close to 1\,M$_\odot$\,yr$^{-1}$ in the case of $\eta$\,Carinae-like eruptions \citep{Smith_2011, Smith_2017_review}. This mass loss produces dense circumstellar material (CSM) proximate to the star which may have a mass exceeding 10\,M$_\odot$ \citep{smith03}. Outbursts from the LBVs may then interact with this CSM \citep{smith13}, boosting the luminosity of the transient due to efficient energy conversion, and producing relatively narrow spectral features, most obvious in the hydrogen Balmer series. These outbursts therefore mimic the interacting Type IIn SNe \citep[SNe\,IIn;][]{Filippenko_1989, Schlegel_1990, Filippenko_1997, Ransome_2021}, and hence are also often called ``SN impostors'' \citep[e.g., SN\,1961V, SN\,2000ch, SN\,2002kg, and SN\,2016blu;][]{VanDyk_2000,Weis_2005,Smith_2011,Kochanek_2011,VanDyk_2012, VanDyk_2013, ws22, Aghakhanloo_2023, Aghakhanloo_2023_16blu,Muller_2023, Aghakhanloo_2025}.

In some cases, SN impostors precede true SNe. The most notable example is SN\,2009ip, which had an initial brightening to $M_V \approx-14.5$\,mag \citep[][]{2009ip, Miller_2009, Berger_2009, Smith_2010_09ip} before exploding as an SN\,IIn in 2012 after a period of fitful activity \citep[][]{sm12,mauerhan13,mauerhan14,Prieto_2012, Graham14, Smith13_09ip,Smith_2014_2009ip, Margutti_2014, Thoene_2015}. Precursor emission is relatively common for SNe\,IIn \citep[around one third of SNe\,IIn show some precursor activity;][]{Strotjohann_2020}, with LBVs being a proposed progenitor path of those SNe \citep[e.g.,][]{so06, smith07gy, smith08tf, Gal-Yam_2007, GalYam_2009, Miller_2009, smith11jl, mauerhan13, smith14, Thone_2017, Smith_2017_review, Jencson_2022, Ransome_2024}.

Owing to their relatively low luminosity (compared to SNe) and possible stochasticity/quasiperiodicity, LBV eruptions can currently be observed only in nearby galaxies, and are difficult to detect in bright star-forming regions. The commencement of LSST will facilitate the earlier and deeper detections of all transient types, including LBV eruptions and SN precursors. Early detection and close monitoring of such sources require complementary surveys to supplement the 3-day LSST cadence. In this paper, we present the Shadow survey  (Section\,\ref{sec:Shadow}), which aims to complement LSST to find faint and young transients.  Section\,\ref{sec:data} introduces AT\,2017des, a restless LBV in the Virgo cluster that was detected by Shadow; it serves as an example of the discovery power of the survey. We also describe the photometry and follow-up spectroscopy collected for this object, and Section\,\ref{sec:results} presents the results. In Section\,\ref{sec:lbv}, we compare AT\,2017des with other LBV eruptions, including SN\,2009ip, and discuss the opportunities offered by LSST. We conclude and summarize our findings in Section\,\ref{sec:conc}.

\section{DECam Shadow -- Leveraging LSST for Early Transient Discovery} \label{sec:Shadow}

\begin{figure}
    \centering
    \includegraphics[width=0.95\columnwidth]{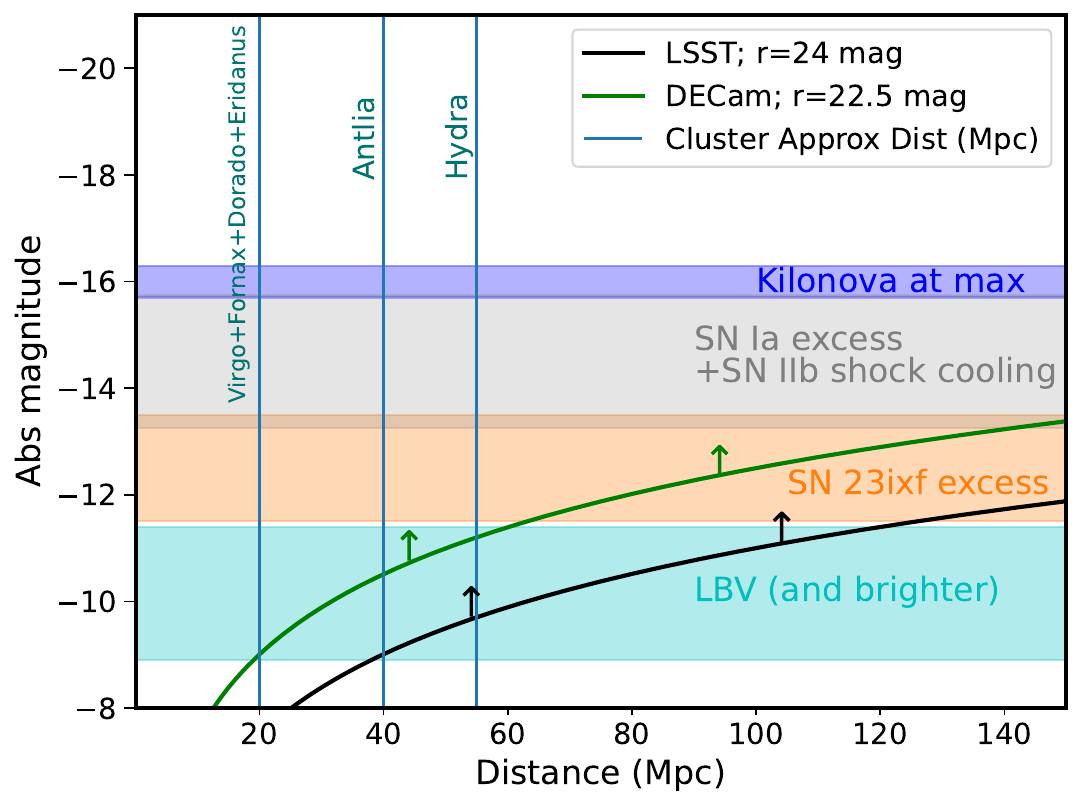}
    \caption{The depths that can be reached by LSST and DECam, demonstrating that early discoveries from LSST can be capitalized upon by the Shadow observing strategy. The blue lines show the approximate representative distances of the clusters that Shadow is targeting, which are outlined in Table\,\ref{tab:fields}.}
    \label{fig:surveydepth}
\end{figure}

%\bf JCW clarify the blue lines in the inset legend. Are those the vertical lines? What clusters?

\begin{figure*}
    \centering
    \includegraphics[width=0.95\textwidth]{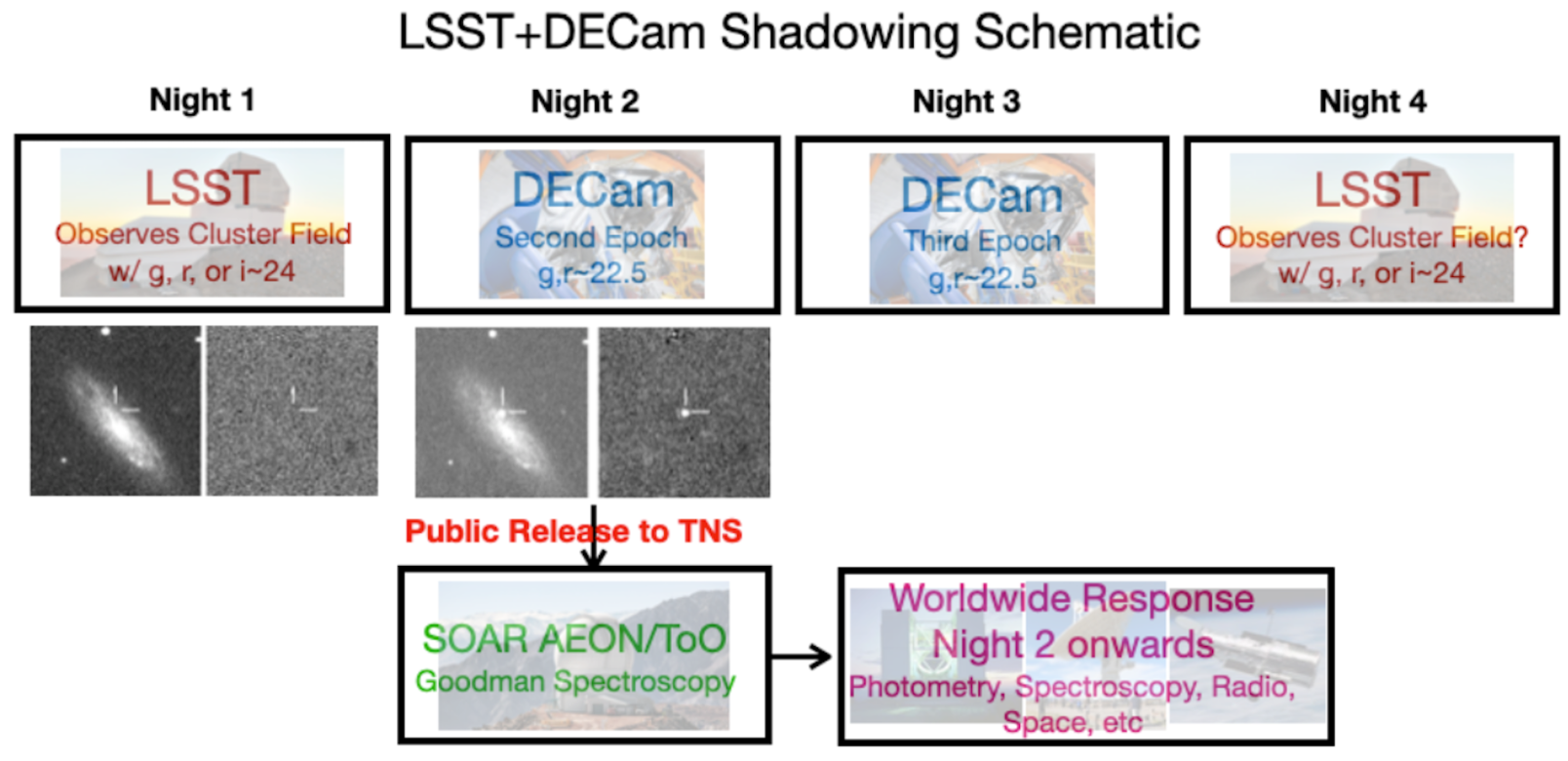}
    \caption{Schematic of the DECam Shadow survey observing strategy, filling in the cadence of LSST. We also show our plans for discovery reporting to the TNS and follow-up spectroscopy. }
    \label{fig:survey}
\end{figure*}

LSST will provide an unprecedented data deluge \citep[with 10\,TB of data produced per night;][]{LSST2}\footnote{\url{https://rubinobservatory.org/for-scientists/rubin-101/key-numbers}}. Nominally, in addition to the Deep Drilling Fields, there will be a 3-day cadence for LSST fields \citep[][]{Bianco_2022}, also known as the Wide Fast Deep survey. Supplementary strategies and surveys are required to maximize the science output of this revolutionary survey. 

A key science target that is facilitated by LSST is the early detection and characterization of transients. The early phases of core-collapse SNe probe the shock breakout and cooling, which provides information on the envelope mass and progenitor radius \citep[e.g.,][]{Soderberg_2008, Modjaz_2009, Modjaz_2019,Hosseinzadeh_2023, Subrayan_2025}, and also the outer layers of the star and confined CSM, probed through light-curve modeling \citep[e.g.,][]{Dessart_2017,Morozova_2017,Morozova_2018}, as well as early interaction signatures in the spectra known as ``flash ionization features'' \citep[e.g.,][]{Quimby_2007,Gal-Yam_2014, Khazov_2016,Yaron_2017, Kochanek_2019, Bruch_2020,Jacobson-Galan_2022, Jacobson-Galan_2023, Bostroem_2023, Bruch_2023,Shrestha_2024_24ggi, Jacobson-Galan_2024, Jacobson-Galan_2024_ggi, Ransome_2026_25ngs}. For thermonuclear SNe, the early phases of the transient may show signs of CSM interaction \citep{Jiang_2021, Srivastav_2023}, and signatures of a companion star in a single-degenerate system being shocked by the SN ejecta \citep[e.g.,][]{Kasen_2010, Hosseinzadeh_2017_iacomp,Hosseinzadeh_2022,Hosseinzadeh_2023bee, Magee_2022}. In addition to this early SN science, LSST will discover gap transients, variable activity, precursor activity, and late-time rebrightening of older transients, which may be as faint as $-9\,$mag out to around 40\,Mpc, as can be seen in Figure\,\ref{fig:surveydepth}.  In conjunction with complementary surveys and other follow-up facilities, LSST discoveries will allow us to build ``gold samples'' of transients with spectral time-series and high-cadence photometry, including those discovered at early times, and unusual, faint, rare, and exotic objects.

\begin{table*}
    %\centering
    \caption{Summary of the Shadow target clusters. We outline the cluster names, field-center coordinates, distances, area on the sky, the number of fields per cluster, and the approximate $r$-band depth.}
    \label{tab:fields}
    
    %\begin{adjustbox}{center}
    \begin{threeparttable}[b]
    \begin{tabular}{l|c|c|c|c|c|c}
        \hline
        Cluster & $\alpha$ (J2000) & $\delta$ (J2000) & Distance (Mpc) & Area (deg$^2$) & Number of Fields & $M_r$ Depth  (mag) \\
        \hline
        Eridanus  & 03:28:00 & -20:44:00 & 25\tnote{1} & 45 & 15 & $-9.5$  \\
        Fornax    & 03:38:00 & -35:27:00 & 21\tnote{2} & 45 & 15 & $-9.1$  \\
        Dorado    & 04:17:00 & -56:06:00 & 20\tnote{3} & 10 &  5 & $-9.0$ \\
        Antlia    & 10:30:00 & -35:20:00 & 40\tnote{4} &  3 &  1 & $-10.5$ \\
        Hydra     & 10:36:00 & -27:31:00 & 52\tnote{5} &  9 &  3 & $-12.0$ \\
        Virgo     & 12:30:00 & +12:20:00 & 18\tnote{6} & 90 & 30 & $-8.8$ \\
        \hline
    \end{tabular}
    \begin{tablenotes}[flushleft]
        \small
        \item $^1$\citet{Omar_2005}; $^2$\citet{Blakeslee_2009}; $^3$\citet{Kourkchi_2017};\\ $^4$\citet{SmithCastelli_2008}; $^5$\citet{LaMarca_2022}; $^6$\citet{Freedman_1994}
    \end{tablenotes}
    \end{threeparttable}
    %\end{adjustbox}
\end{table*}
%As can be seen in Figure\,\ref{fig:survey}, LSST and Shadow complement each other, where (using conservative $r-$band single-visit limiting magnitudes) 

%\textcolor{red}{talk about the luminosity ranges of different target transients}
The single-visit depth of LSST (24--25\,mag, depending on the filter, $ugrizy$) will produce the earliest detections of transients to date. We aim to capitalize on this capability with the Shadow Survey (PI David Sand, program ID 2025B-589422). Shadow uses DECam \citep{decam, decam2} on the 4\,m Blanco telescope at the Cerro Tololo Inter-American Observatory in Chile, only 10\,km from the Vera C. Rubin Observatory. Our observing strategy is summarized in Figure\,\ref{fig:survey}. The 3-day LSST Wide Fast Deep cadence is not ideal for the monitoring of young, fast-evolving transients, even if found at unprecedented early times. Shadow therefore follows a set of fields that cover six local galaxy clusters after LSST observations.  This strategy fills in the gaps in the cadence by observing the two nights between LSST visits in the $g$ and $r$ filters with 90\,s exposures per filter. Our observing strategy will result in an effective daily cadence of Shadow fields when combined with LSST. As the Shadow fields cover local galaxy clusters, they are rich with nearby possible hosts. These fields are Antlia, Dorado, Eridanus, Fornax, Hydra, and Virgo (which notably contains M\,87 and M\,49; see Table\,\ref{tab:fields} for further details on the Shadow fields). The Shadow fields are chosen over a range of RAs such that there are pointings available all year. Our ideal transient discovery workflow is thus as follows.

\begin{enumerate}
    \item \textbf{Night 1:} LSST observes a Shadow field yielding a deep nondetection.
    \item \textbf{Night 2:} A transient in the Shadow field is now detectable by DECam (22--23\,mag). DECam observes the field via either the DECam Alliance for Transients (DECAT) queue or a target-of-opportunity (ToO) trigger. Our same-day ToO triggers are schedule dependent, but we will likely require $\sim3$--4 per week, scheduled when DECAT is not observing. The Shadow data are processed through the pipeline described in Section\,\ref{sec:pipe}, which finds a new transient after producing difference images. The discovery is reported to the public Transient Name Server (TNS\footnote{\url{https://www.wis-tns.org}}). We then utilize spectroscopic follow-up resources available to Shadow, such as SOAR via The Public AEON Spectroscopic Survey for Transient Astronomy \citep[PASSTA;][]{PASSTA}. SOAR follow-up observations can be made on the same night owing to the proximity to the Vera C. Rubin Observatory. Other global facilities may respond after the public discovery alert. At this point, the transient will be spectroscopically classified. Shadow data are ingested into the Searches After Gravitational-waves Using ARizona Observatories (SAGUARO) target and observation manager \citep[TOM;][]{saguarotom}, where candidate pages show photometry aggregated from sources including LSST, difference imaging, spectra, and crossmatched information from other brokers and the TNS.
    \item \textbf{Night 3:} The second DECam visit, probing the early evolution of the transient. A day 2 spectrum will be obtained through PASSTA and other facilities to which Shadow has access,  including the Las Cumbres Observatory, Lick Observatory, MMT, the Hobby-Eberly Telescope, Gemini, the Southern African Large Telescope, Keck, the Large Binocular Telescope, the Bok Telescope, the Australian National University 2.3\,m, Magellan, and the Konkoly Observatory (which has a dedicated program to follow Shadow candidates).
    \item \textbf{Night 4:} LSST revisits the Shadow field, closing the Shadow loop. Our observations then repeat after this visit, producing an effective nightly cadence per Shadow field. In conjunction with our worldwide follow-up campaign enabled by our access to facilities all over the planet, we will produce exquisitely sampled multiband light curves and spectral datasets for Shadow transients. 
\end{enumerate}

 Our observing strategy and the pipeline described in Section\,\ref{sec:pipe} have been tested during pre-LSST operations \citep[the required LSST templates are being produced as part of the Rubin Early Science program; see][]{RTN-011} by observing Shadow fields through DECAT and ToO triggers. In addition to AT\,2017des which we describe below, Shadow has discovered numerous transients\footnote{An up-to-date list can be seen on the \href{\tnsurl}{TNS}.}, with 88 discoveries reported at the time of writing (after observations over $\sim60$ nights). These transients include four (background) SNe that have been spectroscopically classified after the Shadow discovery, including SN\,2026ftl \citep[SN\,Ia;][]{SN2026ftl_disc, SN2026ftl_class}, SN\,2026guf \citep[SN\,Ia;][]{SN2026guf_disc, SN2026guf_class}, SN\,2026ixd \citep[SN\,II;][]{SN2026ixd_disc, SN2026ixd_class}, and SN\,2026jmu \citep[SN\,Ia;][]{SN2026jmu_disc, SN2026jmu_class}. We have also discovered possible classical novae in M\,49 \citep[e.g., AT\,2026ftn;][]{AT2026ftn_disc} and M\,87 \citep[e.g., AT\,2026hvu;][]{AT2026hvu_disc}. In addition to SNe, classical novae, and massive stellar outbursts, Shadow may also discover other rarer and exotic members of the ever-growing transient zoo such as luminous fast blue optical transients \citep[][]{Perley_2019_cow, Ho_2023}, ambiguous nuclear transients \citep[][]{Hinkle_2022,Subrayan_2023}, and gamma-ray burst/orphan afterglows \citep[][]{Sari_1998, Woosley_2006, Lamb_2018, Perley_2025}, offering the unprecedented opportunity for early discovery and daily-cadence observations of these more elusive classes (albeit the rates of such objects will be low in our fixed fields). We now outline the pipeline that Shadow uses to detect transients and generate discovery alerts.

\subsection{The Shadow Pipeline} \label{sec:pipe}

The first step in the Shadow pipeline is to generate the observing plan for the night. We do this by querying the LSST alerts \citep[][]{lsst_prompt,lsst_alerts} from the previous night with a custom Shadow filter on ANTARES\footnote{\url{https://antares.noirlab.edu/}} \citep{antares}, which crossmatches the alerts with our list of Shadow fields to assess overlap. Any matching fields are added to the Shadow schedule. We then query the publicly accessible Rubin Schedule Viewer service for the night\footnote{\url{https://usdf-rsp.slac.stanford.edu/obsloctap/static/viewer.html}, \url{https://usdf-rsp.slac.stanford.edu/obsloctap/skymap}}. If a Shadow field is being observed by LSST, we do not add it to the Shadow plan, and do add fields that are not coincident with LSST pointings. If there are no LSST alerts from the previous night, and no Shadow fields in the LSST schedule, all of those fields (that are visible) are added to the Shadow schedule. The final target list is then submitted to either DECAT or as a ToO, depending on the DECam schedule\footnote{At the time of writing, LSST has not yet begun in earnest, and the Rubin Early Science Program is in the process of making templates for LSST fields; therefore, our observing plan has been focused on the development of project infrastructure.}. 

After each Shadow observing night, our DECam $g,r$-band data are processed by the NOIRLab DECam Community Pipeline \citep{decamcompipe}, which performs initial reduction of the images (e.g., biasing, flat-fielding, astrometric solution, and background flattening) and is made available on the NOIRLab Astro Data Archive. These data are then fed into a modified version of the data pipeline developed by \citet{Hu_2026}, which has also been used for gravitational-wave counterpart searches, identifying high-redshift transients, and other discoveries \citep[e.g.,][]{Cabrera_2024,Hu_2025,Hall_2026_ulz, Hall_2026_12cm, OConnor_2026}. Our modifications to this pipeline primarily concern the conversion from a GPU-based protocol to CPU, and changing the forced-photometry procedure for our use case, such that the pipeline does not automatically perform forced photometry on all previous epochs for every new candidate so that the pipeline runs as fast as possible. The pipeline automatically queries the archive (based on observation date) and downloads the data to a networked machine. The pipeline sky-subtracts our images, and then makes a template image from either deep pre-existing datasets  or our own images. The difference images are produced using \texttt{SFFT} \citep[][]{Hu_2022}. \texttt{SExtractor} \citep[][]{1996A&AS..117..393B} finds sources in the difference images which are then run through a convolutional neural network (CNN) designed for DECam time-domain data \citep[][]{Hu_2026} based on similar work \citep[][]{He_2015, Cabrera-Vives_2017, Sun_2022_cnn} to find the most probable real sources via a real-bogus score. The pipeline crossmatches with variable-star catalogs (also requiring a detection in both filters to rule out moving objects) before producing a final list of candidates \citep[a detailed description of the pipeline is given by][]{Hu_2026}.

The final candidate list is then automatically ingested into the SAGUARO TOM. The candidates are crossmatched with additional catalogs, such as list of active galactic nuclei \citep[AGNs;][]{flesch23_milliquasv8}, the Minor Planet Center, and other variable-star catalogs\footnote{Including ASAS-SN variable-star catalog X \citep{shappee_man_2014, christy_asas-sn_2023}, Pan-STARRS variable-star catalog \citep[][]{Beck+21}, and Gaia \citep[][]{gaia_collaboration_gaia_2023}.} \citep[as described by ][]{Rastinejad22,Franz25}. SAGUARO also associates the transients with the most probable host galaxies using a probability of chance coincidence metric \citep[see][for more details]{saguarotom}\footnote{These galaxy catalogs include GWGC  \citep[Gravitational Wave Galaxy Catalog;][]{White+11_GWGC}, SDSS DR12 (Sloan Digital Sky Survey Data Release 12) photo-$z$ catalog \citep{alam_eleventh_2015}, GLADE \citep{dalya_glade_2022}, Hecate \citep{Kovlakas+21_HECATE}, Pan-STARRS (PS1) Galaxy Catalog \citep{Beck+21}, Legacy Survey Data Release 10  \citep[LS DR10;][]{Zhou+23_LSDR10}, and DESI (Dark Energy Spectroscopic Instrument) Early Data Release \citep{desi_collaboration_early_2024}.}. The candidate pages display image triplets showing the science, template, and difference images, as well as the machine-learning score from the CNN real-bogus classification and the brightness of the object. A light curve is also presented, with our DECam data and any other aggregated data, such as from the Asteroid Terrestrial-impact Last Alert System (ATLAS) Forced Photometry service \citep{ATLAS}. Through SAGUARO, we can report a newly discovered transient to the TNS. Follow-up observations can be scheduled on SAGUARO automatically using API wrappers such as PyMMT \citep[][]{pymmt,Shrestha_2024_23axu}. Any resultant spectroscopic data may also be uploaded to SAGUARO. Candidates that have been vetted and followed up using our pipeline and SAGUARO can then be analyzed, characterized in more detail, and added to transient samples. In the next section, we describe the LBV outburst of AT2017des, a transient detected with the Shadow Survey.

\section{The Restless LBV AT\,2017des} \label{sec:data}

\subsection{Discovery and Characterization}

Initially discovered on 2017-04-16\footnote{UTC dates are used throughout this paper unless otherwise indicated.} by Pan-STARRS \citep[][]{PS1} with the internal name PS17cke \citep[][]{2017des_disc}, AT\,2017des ($\alpha = 12^{\rm hr}34^{\rm m}18.89^{\rm s}$, $\delta = +06^\circ 28' 26.94''$; J2000) was spectroscopically classified as an LBV\footnote{After being identified as a contaminant in gravitational-wave counterpart searches \citep[][]{McBrien_2021, Fulton_2025}.} by \citet{2017des_class} after previous attempts to classify the transient were inconclusive \citep[][]{2017des_1st_class}. AT\,2017des resides in the Virgo-Cluster galaxy NGC\,4532, at a redshift of 0.0067 \citep[we assume a distance of 15\,Mpc, $v=2061$\,km\,s$^{-1}$, and $\mu=30.9$\,mag to be consistent with the Virgo, Great Attractor, and subcluster-corrected values presented by][]{McBrien_2021}. AT\,2017des was also detected by ATLAS in 2021, with the internal name ATLAS21noy, and by ZTF in 2026 \citep{ztf} with the name ZTF26aaufnkd. 

\begin{figure}
    \centering
    \includegraphics[width=0.95\columnwidth]{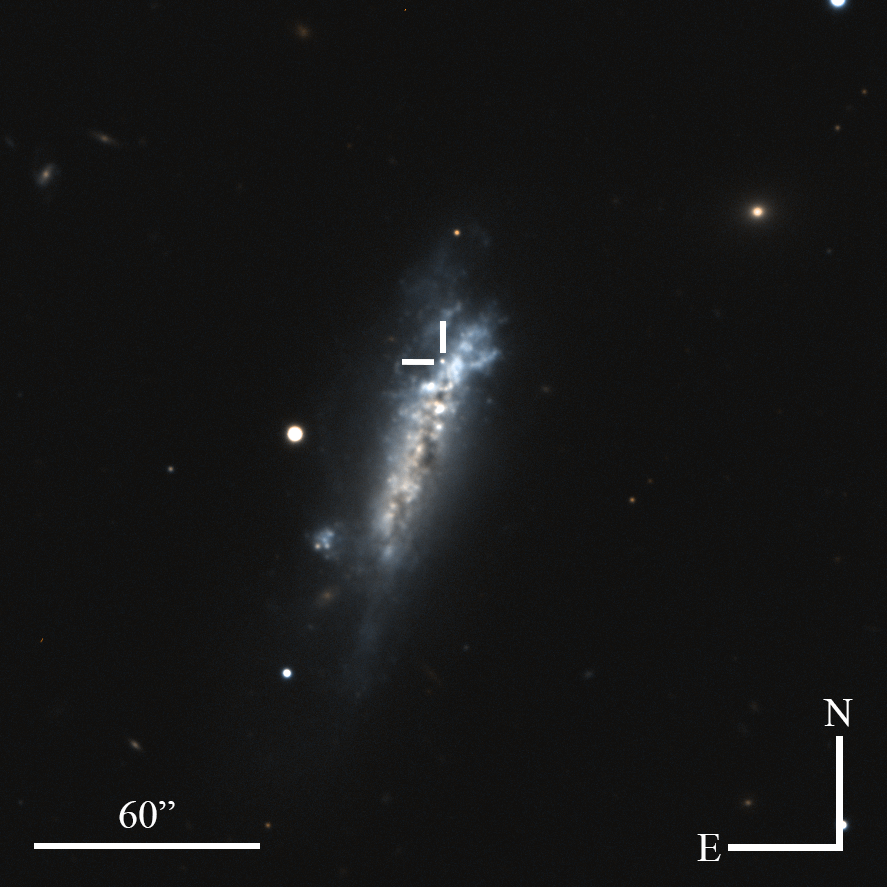}
    \caption{Color-composite image of AT\,2017des in its host, NGC\,4532, using our DECam $g$ and $r$ images. These images were obtained on 2026-04-06.}
    \label{fig:prettypic}
\end{figure}

\begin{figure}
    \centering
    \includegraphics[width=1\columnwidth]{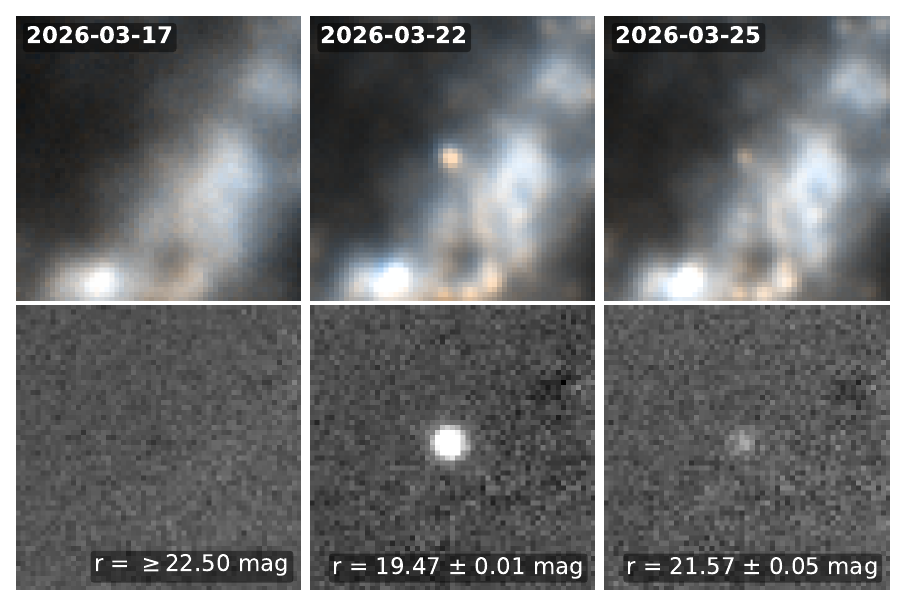}
    \caption{\textit{Top:} Color-composite cutouts of AT\,2017des using our DECam $g$ and $r$ images at three different epochs within an 8 day timespan, along with the corresponding $r$-band difference images ({\it bottom}). These data show the rapid variability of AT\,2017des. All cutouts are 15$^{\prime\prime}$ squares centered on AT\,2017des, and are oriented with north being up and east to the left.}
    \label{fig:diffs}
\end{figure}

AT\,2017des was detected by the DECam Shadow survey on 2026-02-14 at $20.22\pm0.01$\,mag in the $r$ band (single-filter detection, with the first $g$- and $r$-band detection being on 2026-03-22). Thereafter we initiated follow-up observations, noticing a relatively rapid evolution. A DECam image of AT\,2017des within its host is presented in Figure\,\ref{fig:prettypic}. We also show a series of difference images and corresponding color composites of AT\,2017des, displaying the rapid variability of this transient, in Figure\,\ref{fig:diffs}.

\subsection{Photometry} \label{sec:phot}

We have 11 photometric epochs in the range MJD\,61085.2--61165.1 collected using DECam in the $g$ and $r$ bands. The Pan-STARRS data are taken from SAGUARO after they were ingested from TNS \citep[the initial discovery photometry at MJD\,57859;][]{2017des_disc}. We also collect $g$- and $r$-band data from the ZTF forced-photometry service \citep[][]{Masci_2019, Masci_2023}. The ZTF data cover a baseline of  $\sim3000$ days. We supplement these data with imaging from Binospec \citep[][]{Fabricant_2019_binospec} on the MMT in the $r$ and $i$ bands. These MMT observations were reduced using the \texttt{POTPyRI}\footnote{\url{https://github.com/CIERA-Transients/POTPyRI}} reduction pipeline  which performs flat-field, bias, and dark-frame corrections, solves the World Coordinate System (WCS) using stars in the Gaia DR3 catalog, and stacks the individual exposures. The images were then zeropoint calibrated using stars in Pan-STARRS Data Release 1 \citep{Chambers_2016}. Point-spread-function (PSF) photometry of AT~2017des was done using a PSF model built from bright isolated stars in the image. 

Further observations were undertaken by the Las Cumbres Observatory \citep[LCO;][]{lco} as part of the Global Supernova Project \citep[GSP;][]{gsp}, which uses a global network of robotic telescopes. These data employ a $griz$ filter set. We processed the LCO data using \texttt{lcogtsnpipe} \citep[see][for details]{Valenti_2016}.
Additional photometry was collected with the 0.6/0.9\,m Schmidt telescope equipped with a 10k$\times$10k STA 1600L CCD and Pan-STARRS $w$-band filter at Piszk\'estet{\H o} station of Konkoly Observatory, Hungary. Aperture photometry of AT~2017des was computed via difference imaging, tied to the PS1 $r$-band magnitudes of nearby comparison stars.
For our photometric dataset, we take $E(B-V)=0.021\,$mag from \texttt{dustmaps} \citep[][]{Green_2018, Schlafly_2011} and apply the standard reddening law of \citet{Cardelli_1989}, using $R_V=3.1$.

Most of our photometry is from difference imaging. As AT\,2017des is highly variable, the templates likely contain some flux from the transient. To estimate the flux offset from our DECam light curve, we use our observations from MJD\,61137 and MJD\,61138. We have a $g$- and $r$-band detection on MJD\,61137, and only an $r$ detection on MJD\,61138. The $g$ limit is $\sim22.5\,$mag for our 90\,s exposures, and the $g-r$ colors where we have detections in both filters are $\sim1\,$mag. The $r$ detection on MJD\,61138 was at $21.39\pm0.04$\,mag. Therefore, AT\,2017des would be around the limiting magnitude for the $g$ band, indicating a small offset on the order $1\,\mu$Jy. This indicates that our template from MJD\,61082 contained the transient, but at a relative trough in its variability. As the estimated offset for our DECam data is small, we do not adjust our photometry. The ZTF data, however, use templates from 2018. As there are contemporaneous detections between DECam and ZTF data, we can calculate an offset of $\sim0.1$\,mag over observations within 0.2 days of each other. All of the ZTF photometry is shifted to reflect this offset. For the cases where we get a bright $r$-band detection but no $g$-band detection, there are image-subtraction artifacts affecting our photometry owing to the complex embedded environment of AT\,2017des\footnote{At the time of writing, our pipeline is optimized for gravitational-wave counterpart searches rather than for detecting transients in nearby, extended, strongly star-forming hosts. An updated version of our pipeline will be implemented in the near future, but these early results show the power of the Shadow survey regardless.}.

\subsection{Spectroscopy} \label{sec:spec}

After detecting AT\,2017des, we initiated optical spectroscopy and obtained archival data. 
%After the detection of AT\,2017des, we initiated optical spectral follow-up, as well as collecting archival data. 
Our spectroscopic dataset and sources include the following.

\begin{enumerate}
    \item \textbf{EFOSC2:} The first spectrum from 2017-04-22 after the initial Pan-STARRS discovery was taken using EFOSC2 on the 3.58\,m New Technology Telescope. This relatively low-resolution spectrum mostly shows emission from the host, hence the inconclusive classification report by \citet{2017des_1st_class}. We downloaded this spectrum from TNS.
    \item \textbf{Blue Channel:} One spectrum was obtained with the Blue Channel spectrograph \citep{Angel_1979,Schmidt_1989} on the 6.5~m MMT on 2026-04-07. We used a $1.0\arcsec$-wide slit with the 1200 lines mm$^{-1}$ grating ($R\approx3340$) covering a range of $\sim5700$--7000\,\AA. Standard reduction 
    %for the MMT/Blue Channel observation 
    was carried out using {\tt IRAF} (\citealt{Tody_1986}), including bias subtraction, flat-fielding, and optical extraction of the spectrum. Flux calibration was achieved using a spectrophotometric standard star observed at an airmass similar to each science frame, and the resulting spectra were median combined into a single one-dimensional (1D) spectrum.  
    \item \textbf{Binospec:} We also acquired a spectrum using Binospec \citep[][]{Fabricant_2019_binospec} on the MMT, with the G1000 grating which covers $\sim5750$--7200\,\AA. It was taken on 2026-04-16 and was reduced using a standard \texttt{IDL}-based pipeline \citep[][]{Kansky_2019}.
    \item \textbf{LRS2:} We also have two spectra from 2026-04-22 (red and blue arms) and spectra split between 2026-06-10 and 2026-06-12 (blue and red arms, respectively) from LRS2 \citep[][]{lrs2_2,lrs2} on the Hobby-Eberly Telescope. These data were reduced using the \texttt{Panacea} package\footnote{\url{https://github.com/grzeimann/Panacea}}.
    \item \textbf{GMOS-S:} We have a Gemini-S spectrum from 2026-04-26. It was taken with GMOS (B480/R400 gratings with $1''$ slit width) as part of a SN precursor program to classify low-luminosity transients discovered by ZTF and/or LSST (GS-2026A-Q-122; PI W. Jacobson-Gal\'an). The spectrum was obtained because AT\,2017des had been reported as AT\,2026kqq/ZTF26aaufnkd \citep[][]{Sollerman26, 2026kqq}. 
\end{enumerate}

\section{Results} \label{sec:results}

\subsection{The Photometric Properties of AT\,2017des}

\begin{figure*}[!h]
    \centering
    \includegraphics[width=0.95\textwidth]{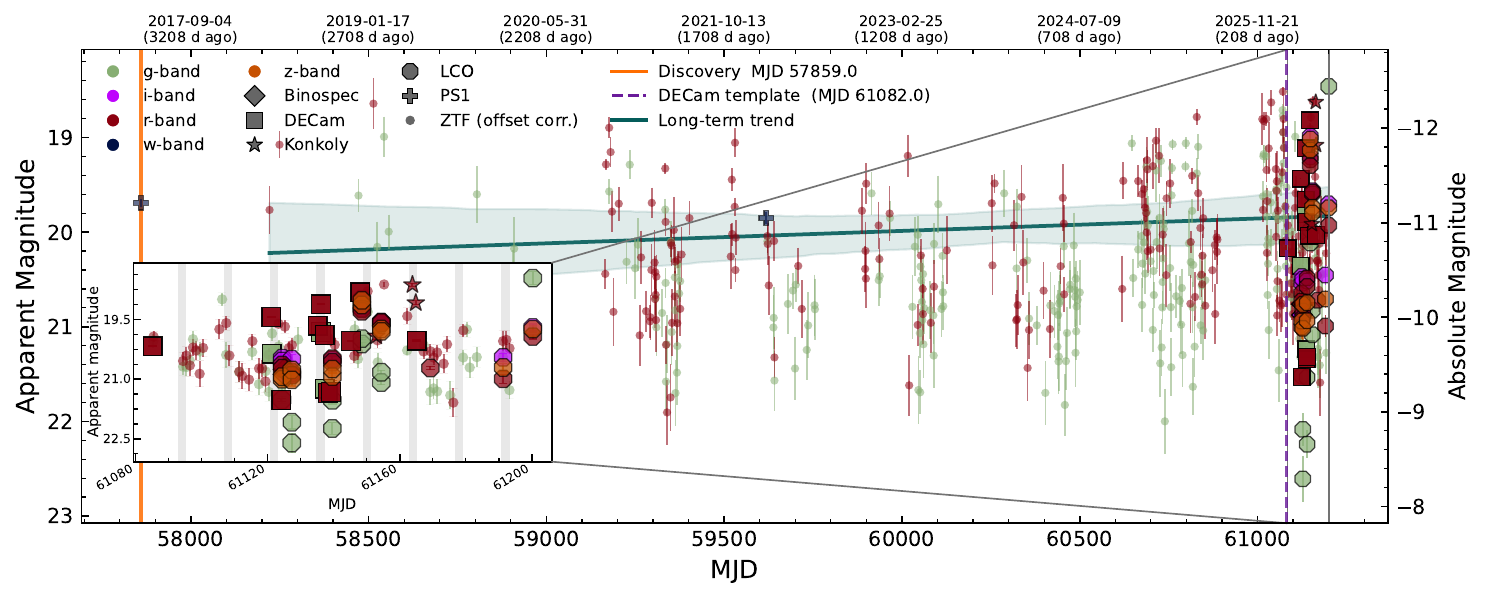}
    \caption{The combined light curve of AT\,2017des with DECam and ZTF $gr$, LCO $griz$, Konkoly $r$, and Binospec $ri$ data. We show the full 9\,yr baseline, starting with the initial PS1 discovery (orange vertical line), with a zoom-in on the recent epoch covered by Shadow. 
    %In this most recent epoch shown in the inset, we 
    This inset displays a rough estimate of the quasiperiodicity seen in AT\,2017des with gray bars spaced out by 14 days and 2 days thick. We indicate the gradual increase in brightness as a blue line with the error region from the fit shaded. The DECam template date is denoted by a dashed purple line.\footnote{Data behind this figure and others in this work can be found on Zenodo at \url{https://zenodo.org/uploads/20752952}.}}
    \label{fig:lc}
\end{figure*}

\begin{figure*}[!h]
    \centering
    \includegraphics[width=\textwidth]{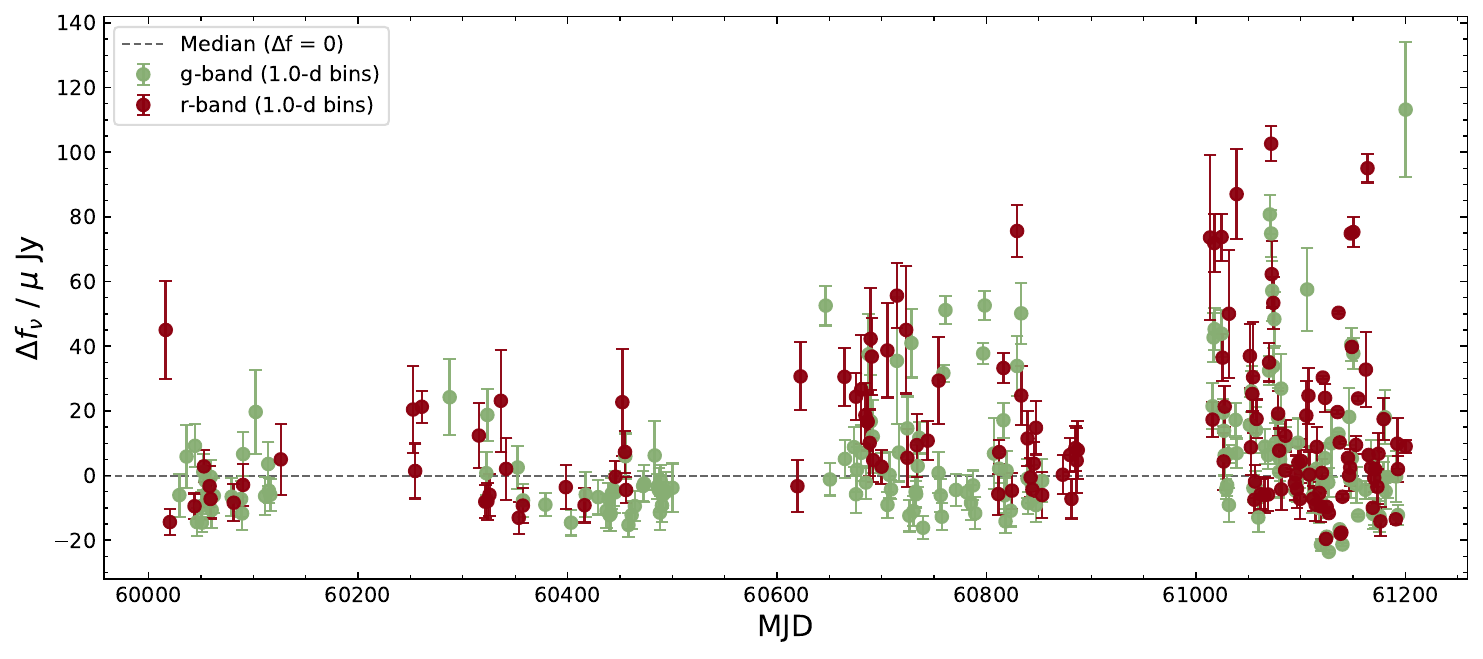}
    \caption{The $g$- and $r$-band light curves of AT\,2017des in flux space relative to the median flux level of the transient in daily bins (with all photometry sources combined). We truncate the date range of this light curve to where most of the ZTF data were taken. The outbursts are getting brighter over time.}
    \label{fig:deltaflux}
\end{figure*}

We present our 3302-day baseline light curve of AT\,2017des in Figure\,\ref{fig:lc}. The object  varies between $-9.0\,$mag and $-12.5$\,mag, with the amplitude of the peaks increasing over time (with the flux of the peaks being around 4--5 times higher at MJD\,61200 than at MJD\,60000). We fit a line with a Markov chain Monte-Carlo (MCMC) routine to the $r$-band data in order to estimate any possible brightening, finding an average increase in brightness of  $\sim 0.05\pm0.01$\,mag per year in the $r$ band over the baseline starting from the first ZTF data ($\sim$MJD\,58250). There may also be some increase in this rate at later times, after $\sim$MJD\,60000. There is no observable corresponding brightening of the trough level, which fades to detection limits. In Figure\,\ref{fig:deltaflux}, we show the light curve (with daily bins) in terms of a change in flux from a median level. We again see that the outbursts are brightening over time relative to the median flux, with the most recent epochs reaching $-12.5$\,mag in $r$ leading to a $\sim3.5$\,mag range in brightness during the most recent epochs. 

The brightness of AT\,2017des changes rapidly. For example, there is a 0.8\,mag fading between observations one day apart (MJD\,61,137 and MJD\,61,138). These rapid variations are not representative of typical S\,Doradus episodes or LBV giant eruptions, both of which typically last for years or decades \citep{smith26}. Instead, the rapid flickering of AT~2017des is reminiscent of objects like SN~2000ch \citep{Aghakhanloo_2023}, AT~2016blu \citep{Aghakhanloo_2023_16blu}, brief precursor flares in the 19th century eruption of $\eta$\,Carinae \citep{Smith_2011_etacar}, and the progenitor of SN~2009ip \citep{Smith_2010}. There have been indications that some such LBV eruptions are periodic \citep[e.g.,][]{Smith_2011_binlbv,Aghakhanloo_2023, Aghakhanloo_2023_16blu}, which have been suggested to be linked to interactions with an eccentric binary. We therefore probe for periodic behavior using a Lomb-Scargle algorithm. We do not find strong evidence for a steady period to the outbursts, suggesting stochasticity or quasiperiodicity. Note, however, that from the light curve and our observations, the intervals between outburst peaks are on the order 10\,days. We highlight this in the inset in Figure\,\ref{fig:lc} with gray bands. The periodicity from periastron companion passages is $\sim100$\,days. As the variability of AT\,2017des is an order of magnitude shorter than this and not clearly periodic, AT\,2017des may have a closer companion in a more circular orbit. The variability is then caused by some instability in the system, such as within an accretion disk.

% \N{(perhaps worth noting that you would only expect periodicity in longer-period (like morethan 100 d) and significantly eccentric system.  this is because the increase in brightness arises when the stars come closer together at periastron, increasing the accretion rate or whatever interaction is making them brighter.  in a shorter period with a circularized orbit, you won't see periodicity in the light curve due to the orbit because the stars are always at the same separation.  in this case, variability would be due to some inherent instability (like in the accretion disk around a compact companion, or whatever... )}

\subsection{Spectroscopic Evolution}

\begin{figure*}[!h]
    \centering
    \includegraphics[width=0.75\textwidth]{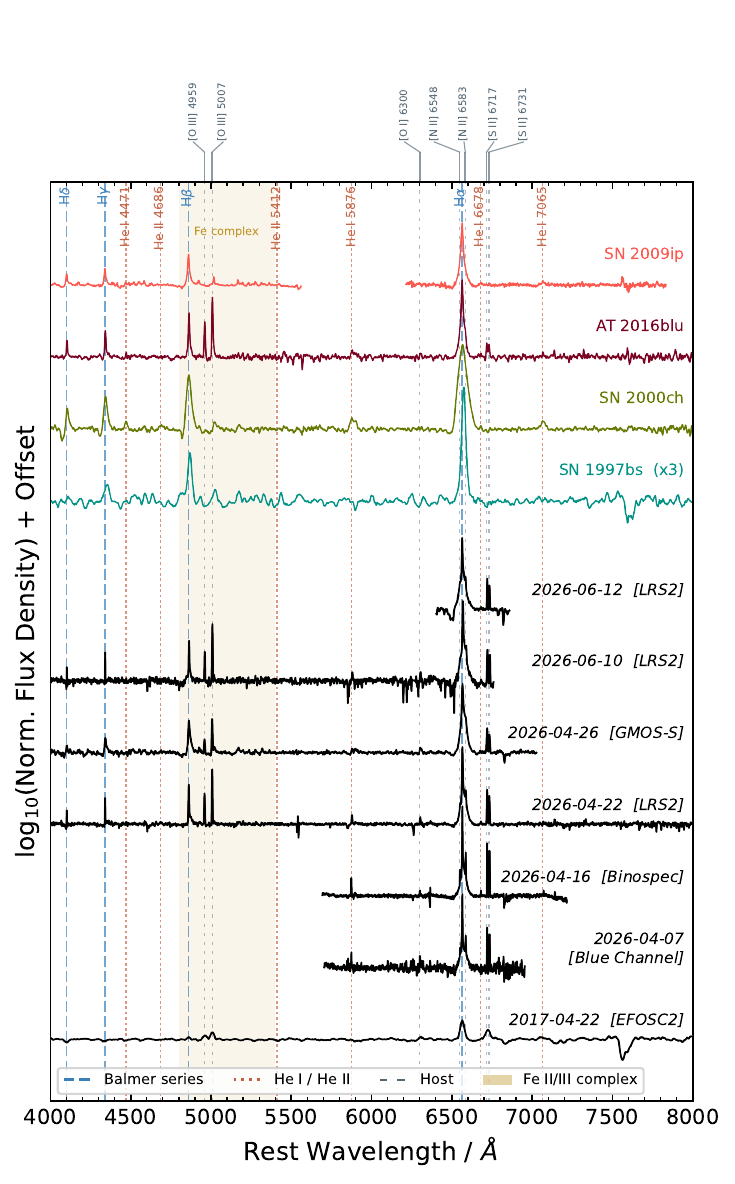}
    \caption{The spectral time-series of AT\,2017des. We record the date each spectrum was taken and by which instrument. Also shown are spectra of comparison objects SN\,1997bs \citep[which is scaled for legibility][]{VanDyk_2000}, SN\,2000ch \citep[][]{Wagner_2004, VanDyk_2013, Aghakhanloo_2023}, AT\,2016blu \citep[][]{Aghakhanloo_2023_16blu, Aghakhanloo_2025}, and pre-2012 SN\,2009ip \citep[][]{Smith_2010_09ip}. We mark host lines and other features such as the Balmer series, and we shade the region where an iron complex is apparent in LBV outbursts.}
    \label{fig:spec}
\end{figure*}

Our spectral time-series of AT\,2017des, and comparisons with other LBV outbursts, is presented in Figure\,\ref{fig:spec}. We note that the spectrum from 2017 only has host-galaxy emission, and that the relatively low-resolution GMOS-N spectrum from 2026-04-26 shows the host [N\,II] lines blended with H$\alpha$. There is strong host emission in the AT\,2017des spectra, consistent with the position of the transient within its host. As is typical with LBV outbursts, the spectrum is dominated by strong hydrogen Balmer series emission, with the lines having Lorentzian profiles; the electron-scattering wings arise from surrounding material, such as CSM, dense winds, or an accretion disk. We also show the region where there is a forest of iron lines seen in SN\,2009ip, SN\,1997bs, and possibly AT\,2016blu. These iron features are apparent in the most recent spectrum of AT\,2017des. While there are multiple instances of the transient fading below our limits over the follow-up period with Shadow and the MMT/HET spectra, indicating fitful episodic outbursts, there are no appreciable differences in the spectra, apart from changes in the H$\alpha$ line strength and P\,Cygni absorption. We note that in our latest GMOS and LRS2 spectra, there is a P\,Cygni profile developing in H$\alpha$ and H$\beta$. This is indicative of absorption from line-of-sight material. While it seems to be a developing profile, as there is over a month between these last spectra, there may not be a causal link in this development, as other outbursts will have occurred in the gap between spectral observations.

% \N{(we don't know this.  you also can get e- scattering wings in a dense wind or in a dense accretion disk.  lorentzian wings do NOT indicate CSM interaction...they indicate dense ionized gas, which could arise from multiple scenarios.)}

%SN\,2002kg \N{(you could probably delete 02kg from this comparison - it is a very different type of object.  sortof a puny little weak nothin of an eruption, and slower.)}  \citep[][]{Schwartz_2003, Weis_2005}

\begin{figure}[!h]
    \centering
    \includegraphics[width=\columnwidth]{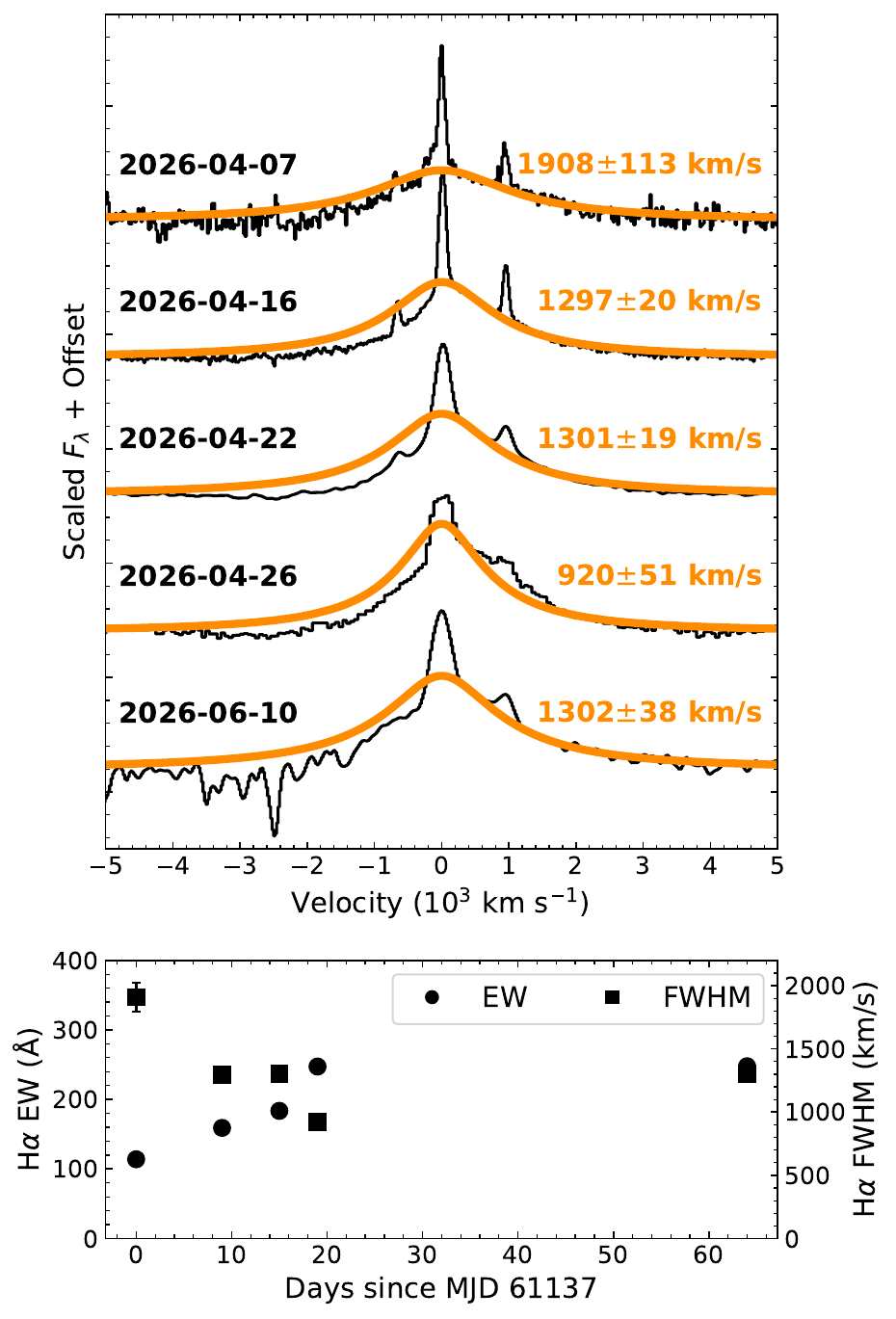}
    \caption{\textit{Top:} Lorentzian fits to the H$\alpha$ profiles in our spectral time-series. We do not include the spectrum from 2017 as it is host-dominated. \textit{Bottom:} The evolution of the H$\alpha$ FWHM and EW from our Lorentzian fits. Here it can be seen that as the FWHM decreases, the EW increases.}
    \label{fig:lorenztian_fit}
\end{figure}

We perform a line-profile fit to the H$\alpha$ emission using a linear background and a Lorentzian function, carried out with {\tt scipy.optimize.curve\_fit}. To fit the broader emission wing and mitigate the narrow emission peak at the center of the H$\alpha$ line profile, we exclude regions within $\pm350$ km s$^{-1}$ and simultaneously fit a narrow Gaussian to the [\ion{N}{2}] $\lambda6583$ emission. Given the slightly asymmetric nature (which may be due to electron scattering and/or asymmetries in the system, discussed in more detail in Section\,\ref{sec:lbv}), we constrain our fit to only the red wing of the H$\alpha$ line profile and force the fit to be centered at 0\,km\,s$^{-1}$ (see Figure~\ref{fig:lorenztian_fit}).

%\N{(if the wings are asymmetric, then they are not necessarily e- scattering wings)}

The equivalent width (EW) of each Lorentzian fit is calculated, as opposed to the EW of the full profile, to minimize effects from host-galaxy emission lines. We trace the evolution of the EW along with the full width at half-maximum intensity (FWHM) of the broad portion of the H$\alpha$ profile. The EW uncertainty is quantified via standard error propagation. The evolution of the FWHM and EW of H$\alpha$ 
%and the EW 
is shown in the bottom panel of Figure~\ref{fig:lorenztian_fit}. We see that over our spectral epochs, the EW of the H$\alpha$ line increases, while the Lorentzian FWHM decreases; such a relation is seen in other objects and is discussed further in Section\,\ref{sec:lbv}.

%\textcolor{red}{LIGHT CURVE IN FLUX SPACE AND LOMB SCARGLE}
%\section{Discussion} \label{sec:discussion}

\section{The Luminous Blue Variable Landscape} \label{sec:lbv}

The eruptions of LBVs are heterogeneous in terms of both their luminosity and variability timescale \citep{smith26}. Many objects discovered in transient surveys are called LBVs that may have little or no similarity to classic LBVs in the Milky Way and Magellanic Clouds. In particular, many SN impostors display brief brightening and fading events that are unlike the decade-long timescales of known LBV eruptions; this may be at least partly because sudden increases in brightness are more readily detected in surveys than gradual year- or decade-long changes. 

\begin{figure}[!h]
    \centering
    \includegraphics[width=\columnwidth]{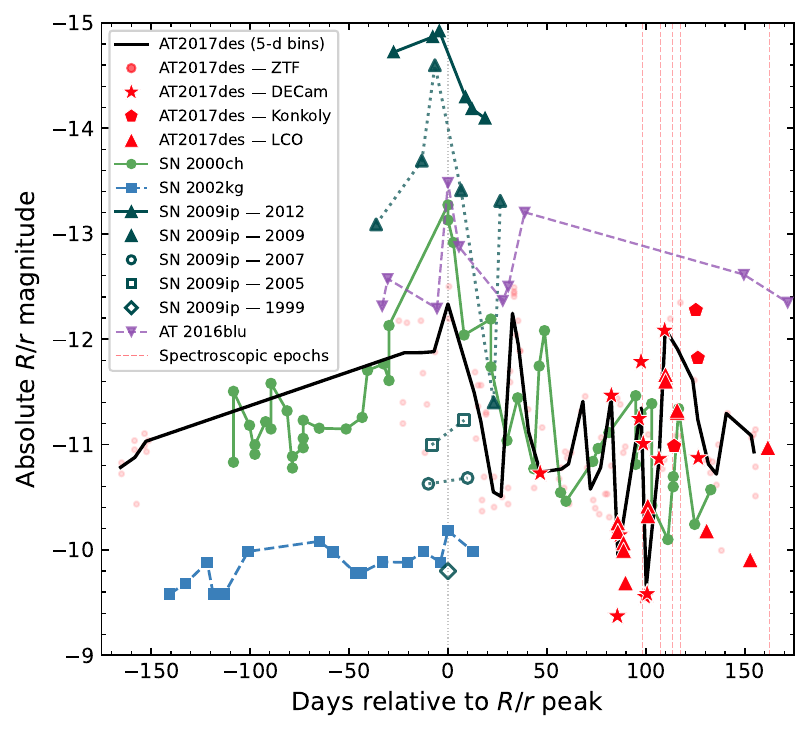}
    \caption{The $r$-band light curve of AT\,2017des compared with other LBV eruptions/SN impostors ($R$ or $r$ band). These comparison objects include SN\,2000ch \citep[][]{Aghakhanloo_2023}, SN\,2002kg \citep[][]{Weis_2005}, SN\,2009ip \citep[][]{Smith_2010,mauerhan13}, and AT\,2016blu \citep[][]{Aghakhanloo_2023_16blu}. The high-cadence observations of the combined ZTF+DECam+LCO+Konkoly dataset for AT\,2017des reveal significant variability over short timespans. We show the different activity levels of SN\,2009ip.}
    \label{fig:lccomp}
\end{figure}

We show the (ZTF+DECam+LCO+Konkoly) $r$-band light curve of AT\,2017des in Figure~\ref{fig:lccomp} compared to other LBV outbursts/SN impostors. Our combined dataset has a significantly higher cadence than the other transients, perhaps revealing more rapid variability; however, some of our companion objects do exhibit rapid variability, such as SN\,2000ch \citep{Wagner_2004, Muller_2023,Aghakhanloo_2023}.  These comparison objects exemplify this heterogeneity, with an $\sim6\,$mag range in luminosity. While AT\,2017des seems to be more variable than some of the comparison objects, the higher cadence (every few days) of the AT\,2017des data means that this may be an observational effect due to both this high cadence and long baseline. Therefore, our observations of AT\,2017des highlight the importance of long-term monitoring of highly variable sources.

%\N{(some other objects do also show very rapid variability, including 00ch - except that the plot here misses the bright brief peak from the wagner paper for some reason... and there are multiple suchg eruptions in mojgan's papers.)}
%\N{(...ok, in case you can't tell, i'm basically trying to find a diplomatic way to say that transient people who classify things as LBVs are mostly full of BS.  in this context, "LBV" is really just a catch-all term for "i don't know what it is, but it is not a real SN" and we shouldn't think that an LBV classification really indicates that it is anything like LBV stars...)}
%(individual LBV outbursts are also diverse, so may resemble each other for given eruptions).

 The photometry presented in Figure\,\ref{fig:lccomp} shows the 2012A outburst of SN\,2009ip which preceded the 2012B eruption, which was likely the terminal SN explosion \citep[][]{Smith_2022}. This 2012A outburst follows after a period of variability spanning from the first 2009 detections \citep[and the pre-2009 detections found retroactively;][]{Smith_2010_09ip}. The outbursts observed for SN\,2009ip span from S\,Doradus variability \citep[brightening at visual wavelengths from quiescence to around $-10$  or $-11\,$mag; ][]{Smith_2010} preceding the larger outbursts which are akin to the brief flares before the Great Eruption of $\eta$\,Carinae, with $V$-band luminosities peaking around $-14.5\,$mag \citep[see, e.g.,][]{Smith_2010,Mauerhan13a}, similar to what is seen in AT\,2017des. This series of activity culminated in the terminal explosion which peaked at $M_V \approx-18$\,mag, forming a transient that interacted strongly with the material expelled during the prior outbursts \citep[e.g.,][]{Mauerhan13a,Smith_2014_2009ip,Margutti_2014}. 
 
 Owing to the unprecedented behavior exhibited by SN\,2009ip, there were suggestions that this ultimate explosion was an even more extreme nonterminal eruption \citep[see, e.g.,][]{Fraser_2013, Pastorello_2013}, particularly as another suggested SN impostor, SN\,1961V \citep[][]{Goodrich_1989, Filippenko_1995,VanDyk_2012} reached similar luminosities \citep[but see][for suggestions that SN\,1961V was a true SN or a pulsational pair-instability SN]{Zwicky_1964, Kochanek_2011, Woosley_2022}. However, late-time observations reveal that SN\,2009ip is fainter than sources seen in the pre-2009 imaging, indicating that it was a terminal explosion \citep[][]{Smith_2022}. Notably, there is a collection of SN\,2009ip-like objects that exhibit a similar ``ramp up'' to the terminal explosion, such as SNe\,2011fh \citep[][]{Pessi_2021}, 2015bh \citep[][]{Elias-Rosa_2016, Thone_2017, Boian_2018, Jencson_2022}, 2016bdu \citep[][]{Pastorello_2017}, 2019zrk
\citep[][]{Fransson_2022, Soker_2022}, 2023ldh \citep[][]{Pastorello_2025}, 2023vbg \citep[][]{Goto_2025}, and 2024hpj \citep[][]{Salmaso_2024}. Some of these SNe have precursor activity that does not reach an $\eta$\,Carinae-like brightness, rather an S\,Doradus-like variability gives way to the SN, as in the case of SN\,2016jbu \citep[][]{Brennan_2022_16jbu}, with the precursor being at a similar luminosity to the outbursts of AT\,2017des at the time of writing. A gradual average increase in brightness accompanied by flaring activity is also seen in AT\,2016blu, which increased in brightness by around 0.5\,mag over 8\,yrs \citep[][]{Aghakhanloo_2023_16blu}. 

Precursor activity may also persist for years, similar to our long-baseline photometry for AT\,2017des, such as SN\,2021qqp \citep[][]{Hiramatsu_2024} and SN\,2023zkd \citep[][]{Gagliano_2025}, which also produce complex SN light-curve shapes, with double peaks interpreted as ejecta shocking massive CSM shells expelled from the progenitor by the prolonged, extreme mass loss which may have exceeded 10$^{-2}$\,M$_\odot$\,yr$^{-1}$.
%\subsection{SN\,2009ip and other Precursors} \label{sec:09ip} \N{(yes... but also accompanied by lots of brief flaring activity...)}

Some LBV eruptions, such as SN\,2002kg, have more modest luminosities and variability modes, and others such as SN\,2000ch are brighter with similar rapid variability. Specifically, SN\,2000ch peaked at $M_V\approx-13.5$\,mag, consistent with AT\,2016blu and the precursor activity of SN\,2009ip \citep[][]{Aghakhanloo_2023}. However, SN\,2000ch also exhibited a much longer periodic behavior in addition to the flaring which we also infer for AT\,2017des, with a period of $\sim 200$ days. SN\,2002kg has a fainter peak than AT\,2017des, with smaller, likely long-period observed variation, consistent with S\,Doradus pulsations \citep[with the LBV being known as V37;][]{Weis_2005}. Such variability is seen in Galactic LBVs (the namesake for this variability mode, S\,Doradus, is in the Large Magellanic Cloud), e.g., AG\,Carinae and HR\,Carinae, and weaker variability seen in P\,Cygni \cite[see][for a recent review of local LBV variability, from strongly active, to dormant]{Spejcher_2025}. This heterogeneity may therefore indicate differences in eruption mechanism, with proposed mechanisms outlined in Section\,\ref{sec:intro}.

This enhanced LBV activity may be analogous to the pre-2009 luminosity of the progenitor of SN\,2009ip and what was seen prior to the Great Eruption of $\eta$\,Carinae \citep[and other transients; see][for a review of the diversity of LBV eruptions]{Smith_2011}. Other LBV outbursts such as SN\,2000ch remain hot, unlike S\,Doradus variables, perhaps as a result of the periastron passage of a companion \citep[][]{Aghakhanloo_2023}. While not apparent in AT\,2017des, S\,Doradus variables can be further enhanced if the star passes the bistability gap at temperatures exceeding 21,000\,K, where wind variability from small changes in the luminosity or stellar radius can increase the mass-loss rates, facilitating the formation of CSM shells \citep[][]{Lamers_1995,Vink_2002,Groh_2011a, Groh_2011b}. The most luminous LBV outbursts, such as the 2009 outburst of SN\,2009ip, the Great Eruption of $\eta$\,Carinae, and other SN impostors/precursors are caused by some similar, debated mechanism such as super-Eddington phases \citep[][]{Owocki_2004} and binary interactions via mergers or close passes by compact companions \citep[e.g.,][]{Soker_2006, Kashi_2010, Kashi_2013, SmithArnett_2014, smith18,Aghakhanloo_2023,Aghakhanloo_2023_16blu, Schneider_2024}. 

%\stred{gravity wave pulsations} \N{(nope.  these only occur 1 yr before core collapse.)} \citep[][]{Shiode_2014}, or even pair-instability pulsations \N{(nope. maybe for 61V, but not for any others...)}  \citep[][]{Woosley_2007, Pastorello_2013}.

For SN precursors, there is often a ``ramp up'' in the luminosity of the pre-SN transient event \citep[e.g., the aforementioned SNe\,2009ip, 2021qqp, and 2023zkd;][]{Smith_2010,Mauerhan_2012,mauerhan13,Smith_2014_2009ip, Hiramatsu_2024, Gagliano_2025}. While it is unclear if this gradual apparent brightening prior to a bone-fide SN is some chain of mechanisms (e.g., S\,Doradus variability\,$\rightarrow$\,bistability phase\,$\rightarrow$\,merger), or if such activity can be explained by the evolution of one of these mechanisms (e.g., inspiralling of a companion before the merger), we see that the outbursts of AT\,2017des are brightening over time (Figure~\ref{fig:deltaflux}). The flux of the peaks at MJD\,61100 are approximately 4--5 times higher than around MJD\,60100. This increasing flux may be symptomatic of the erratic variability apparent in some LBVs, but is also seen in SN precursor activity as this ``ramp up'' prior to the terminal explosion. It is therefore imperative that AT\,2017des and similar transients are monitored frequently.

%\textcolor{red}{COMPARISON TO OTHER LBVs RE: SPECTRA FWHM AND EW EVOLUTION}

%\N{(not really.  certainly S Dor outburst spectra look nothing like SNe IIn, and giant eruptions (eta Car - we only have light echo spectra) look quite different too.  for SN impostors it depends what you mean - sure, they both have H lines narrower than normal SNe II, but they are quite different and pretty easy to distinguish.  LBVs and SN impostors usually have much narrower lines (few hundred km/s) than SNe IIn (1-2 thousand km/s) see Fig 11 in Smith et al. 2011.)}, hence the prevalence of SN impostors \N{(SN impostors are not confused with SNe IIn because of their spectra.  it is just because they get brighter and were found in SN surveys, so people called them SMN impostors.  with the rare exception of weirdos like 61V (which didn't have good spectral coverage during the main 1961 peak), it is pretty easy to tell the diff when you get a spectrum.)}

%\N{(not really - they are usually attributed to strong winds in LBVs and SN impostors, although we admittedly don't really know.)}

The spectra of LBV outbursts and SN impostors exhibit complex H$\alpha$ profiles (and the rest of the Balmer series) that have a narrow core with a FWHM of a few 100\,km\,s$^{-1}$ and broader electron-scattering wings with typical widths of 1000--2000\,km\,s$^{-1}$. These complex profiles are indicative of dense winds, accretion disks, and possibly CSM interaction   \citep[similar to what is seen in SNe\,IIn; e.g.,][]{Filippenko_1989,Schlegel_1990, Filippenko_1997, Pastorello_2013, Ransome_2021}. We show our spectral time-series of AT\,2017des in Figure\,\ref{fig:spec}, with comparisons to other LBV eruptions and SN precursors/impostors. Owing to the location of AT\,2017des within its host, there are strong host lines, but the spectra are dominated by the Balmer series. This is similar to what is seen in all the comparison objects, including SN\,2000ch which also has narrow absorption forming a P\,Cygni feature on the Balmer lines \citep[also seen in SN\,2009ip;][]{Smith_2010_09ip}. The lack of any significant evolution in our spectral time-series is reminiscent of what was seen in AT\,2016blu \citep[with broadly similar widths to AT\,2017des as well;][]{Aghakhanloo_2025}. This somewhat uniform evolution is unlike the S\,Doradus cycle where the spectra may significantly evolve through the pulsations \citep[e.g.,][]{Wesselink_1956,massey2000unprecedented}. Also, AT\,2017des exhibits He\,I $\lambda5876$, similar to SN\,2000ch and AT\,2016blu. This may suggest that the wind has a higher helium abundance than that of other objects. Generally, AT\,2017des is largely similar to the LBVs and SN impostors that have rapid flaring activity and remain hot as they brighten, in contrast to more normal LBVs that cool post-outburst; these hot LBVs include the aforementioned SN\,2000ch, SN\,2009ip precursor, and AT\,2016blu \citep[][]{smith26}.

%These comparison objects, whose variability may be induced by some interaction within a binary are distinct from S\,Doradus variables such as SN\,2002kg which have much longer periods (a few decades) and are often of more modest luminosity.

%\N{(no it is not.  it is similar to the weird sub-class of LBVs/impostors that stay hot as they get brighter and have relatively rapid flaring (sn00ch. 16blu, 09ip precursor, those precursor flares of eta Car (but not its main eruption), maybe MCA-1B, HD5980, etc.), but it is very unlike traditional LBVs, which get cool when they brighten and tend to evolve more slowly.  my advice for this section - rather than talking about how 17des is like LBVs and comparing it to SNe IIn, i would emphasize that it is clearly a new member of a small new subclass of objects including 00ch and 16blu and maybe 09ip's progenitor.  these are lumped together with LBVs because, honestly, we don't know what else to call them, but they are distinct in their behavior from normal/traditional LBVs.  best guess is that this is what you get when you have an LBV-like star in a binary with a compact companion.   see my recent review where i talk about this subclass compared to the rest of the LBVs (Smith 2026) and see the discussion in Mojgan's recent papers on 00ch and 16blu.)}

We show the evolution of the Lorentzian FWHM of the H$\alpha$ profiles of AT\,2017des in Figure\,\ref{fig:lorenztian_fit}. Over the 64 days that our 5 spectral epochs span, the FWHM decreases from $\sim2000\,$km\,s$^{-1}$ to $\sim1100\,$km\,s$^{-1}$. This corresponds to an increase in the EW 
%of these Lorentzian line profiles 
from $\sim100$\,\AA\, to $\sim250$\,\AA. Such an anticorrelation is indicative of evolution of the optical depth, which may be attributed to changes in wind density/speed, changes in an accretion disk, or possibly an expanding shell of ejecta and CSM where the material becomes less optically thick to electron scattering (as the Lorentzian FWHM $\propto\tau^{-1/2}$), and recombination strengthens relative to the continuum, hence increasing the EW of the H$\alpha$ line. This has been seen in similar objects such as AG\,Carinae \citep[][]{Stahl_2001}, P\,Cygni \citep[][]{Richardson_2011}, and MCA\,1B \citep[albeit a Wolf-Rayet star with an LBV-like outburst;][]{Smith_2020}. The profiles are asymmetric, with a more prominent red wing in H$\alpha$ compared to the blue side (and is seen in the other Balmer series lines). This asymmetry is common in LBV eruptions and may be attributed to asymmetries in the outbursts/CSM, and absorption from line-of-sight material \citep[e.g.,][]{Smith_2003_etacarwind,Wagner_2004, Weis_2005, Groh_2009, Aghakhanloo_2025}.

% \N{(probably not CSM interaction - more likley to be wind density/speed changes or changes in an accretion disk.  we don't know that it is CSM interaction.  this is kindof a can of worms here.)}  , alternatively, different parts of the wind are being probed over time, changing the line formation region \N{(but in these objects it is almost certainly not CSM interaction like a SN IIn.  it is dense winds.)}

\subsection{Massive Stellar Eruptions in the Rubin Era} \label{sec:rubin}

Currently, large, high-cadence surveys are limited to relatively nearby hosts and/or particularly luminous eruptions for the discovery of LBVs, SN impostors, and SN precursors. Indeed, AT\,2017des is near the detection limit for ZTF at troughs in its variability. LSST will dramatically  increase the discovery rate of LBV eruptions and other nonterminal stellar eruptions. 

A transient of similar luminosity to AT\,2017des (peaking at  $\sim M_r=-12.5$\,mag) will be detectable by LSST within a much larger volume than what is possible with current surveys such as ZTF. Assuming an $r$-band single-visit limit of 24\,mag \citep[][]{PSTN-054}, the peaks of AT\,2017des would be detectable to $\sim200$\,Mpc. Other stellar eruptions that will be discovered in unprecedented numbers include classical novae, luminous red novae, intermediate-luminosity optical transients, and precursor events to normal SNe\,II, as what was detected for SN\,2020tlf \citep[][]{Jacobson-Galan_2022}. Furthermore, deep limits for precursor activity have been found for other objects such as SN\,2023ixf \citep[][]{Dong_2023, Ransome_2024_23ixf} and SN\,2024ggi \citep[][]{Shrestha_2024_24ggi}, and LSST will be sensitive to possible precursor activity of similar SNe within $\sim10\,$Mpc. The estimated rate for LBV-like precursor events to which LSST will be sensitive (e.g., up to the luminosity of the SN\,2009ip precursors in 2009) is  $\sim 112$\,yr$^{-1}$ out to a median distance of $\sim340$\,Mpc, and the precursors to SNe\,II being $\sim 250$\,yr$^{-1}$ to a distance of $\sim140$\,Mpc \citep[][]{Gagliano_2025_precursor}. Many such objects will be out of the range of our spectroscopic facilities owing to their faintness at larger distances, but photometric classifiers and anomaly-detection algorithms may aid in their identification in the LSST era. Given the stochasticity of LBV eruptions, it is difficult to estimate the rate of LBVs that are in outburst. With this being said, assuming all SN\,IIn precursors are these LBV-like eruptions\footnote{It should be noted that the models of \citet{Matsumoto_2022} and \citet{Tsuna_2024} used by \citet{Gagliano_2025_precursor} do \textit{not} necessarily assume an LBV-like progenitor, but the resulting outbursts would be similar to the LBV-like eruptions and SN impostors.}, then a very conservative lower limit for LSST-detected outbursts is on the order of 100\,yr$^{-1}$, dramatically increasing the current known population.

%, or more than the currently known sample of confirmed (`normal') LBVs in the Local Group \citep[][]{Richardson_2018}. 

%\N{(but LBVs in the local group look NOTHING like the SN IIn precursor eruptions.  they are only similar in the CSM mass that they have ejected... (in other words, the similarity to SN IIn progenitors comes from the large masses of their shells from some past unobserved eruption, not from their current variability, with the possible exception of eta Carinae, which is kindof a stand-out freak and unlike most other LBVs).  apples and oranges here.  i would caution that rates of S Dor outbursts and LBV giant eruptions may have little or nothing to do with rates of SN precursor eruptions.)}

%\N{(need to be careful here - some people have called the 2012a event of 09ip a "precursor".  it is counting those that has led to the comment that they may be common.  but those are more likely to be the beginning of the SN explosion itself... but regardless of interpretation, those things are much more readily detected than precursors like the 2009 flare of 09ip, which are fainter and brief, and therefore harder to catch.  bilinski et al. 2015 looked at this in LOSS data and found none.  Ofek et al. found lots of precursors, but that is because they chose to count brighter/slower things like the 2012a event of 09ip and the early bump of 2010mc as LBVs, not as the SN itself, which may be incorrect.)}
%(that show so-called flash features from interaction with confined CSM)

\section{Conclusions and Summary} \label{sec:conc}

In this work, we have introduced the Shadow Survey which will fill in the cadence gaps of LSST observations in rich local cluster fields and integrate with global facilities to maximize the science output of LSST. We describe our workflow and discovery pipeline. An early science result is highlighted to exemplify our science goals, the re-recovery of the LBV outbursts of AT\,2017des in the Virgo-Cluster galaxy NGC\,4532. This paper can be summarized as follows.

\begin{enumerate}
    \item The Shadow survey uses DECam on the 4\,m Blanco telescope at CTIO to observe local galaxy clusters (Antlia, Dorado ,Eridanus, Hydra, Fornax, and Virgo), filling in the gaps in the 3-day cadence of LSST.
    \item DECam Shadow will facilitate the discovery of faint transients, in particular extremely young SNe, which inform explosion physics and help constrain progenitor properties.
    \item Shadow uses an automated pipeline based on the work presented by \citet{Hu_2026}. We have modified this GPU-based pipeline for our needs, including converting to a CPU-based routine.
    \item Shadow has already discovered numerous SNe and other transients, before  LSST started. One such transient that we detected is AT\,2017des, an outbursting LBV in the Virgo-Cluster galaxy NGC\,4532.
    \item We downloaded ZTF data of AT\,2017des and performed forced photometry. When combined with our Shadow data, the PS1 data, and our Konkoly data, we see an increase in flux over a 9\,yr baseline, similar to some SN impostors and SN\,precursors.
    \item AT\,2017des is more luminous than the S\,Doradus cycles of objects such as SN\,2002kg, and is similar to transients such as SN\,2000ch, but resides between these objects and AT\,2016blu, and the precursor activity of SN\,2009ip in luminosity space. AT\,2017des is most like the ``hot'' LBV-like objects such as SN\,2000ch, the precursors of SN\,2009ip, AT\,2016blu, and the flaring that preceded the Great Eruption of $\eta$\,Carinae.
    \item The outbursts are semistochastic with no clear steady period, but the observed variability is on the order 10\,days. This rapid variability highlights the importance of daily cadence observations. 
    \item We collected 5 spectral epochs using the MMT (Binospec and Bluechannel), Gemini-S (GMOS), and HET (LRS2). We find an anticorrelation between the H$\alpha$ EW and Lorentzian FWHM, an optical-depth evolution effect from changing wind densities or speeds, or possibly an expanding and recombining ejecta/CSM shell.
    \item Generally, the spectra of AT\,2017des are similar to those of some SN precursors such as SN\,2009ip and the subset of LBVs that remain hot post-eruption similar to SN\,2000ch and AT\,2016blu. AT\,2017des may be ``ramping up'' to a terminal SN explosion. It is therefore imperative that this transient be closely monitored.
    
\end{enumerate}

%\N{(not true.  00ch is actually more luminous at peak, but those peaks are not shown in the figure here...check this)} seen in numerous other LBV outbursts \N{(this is most likely a wind density effect, not an expanding ejecta/CSM effect...at least in LBVs.  in principle it could be due to a number of different things.)}

% This evolution suggests evolution of the optical depth, indicative of an expanding CSM/ejecta shell, or that different regions of the wind are being probed.

It is clear that as well as the early discovery of infant SNe, LSST will discover unprecedented numbers of precursors and other nonterminal explosive transients. The Shadow survey strategy will complement the upcoming data deluge, facilitating earlier detection, follow-up observations, and characterization of faint and young transients, maximizing the science potential of this revolutionary survey. 

%% Please use the acknowledgment and contribution environments. This will 
%% be anonomyized when the "anonymous" style option is used. 
\begin{acknowledgments}

Time-domain research by the University of Arizona
team and D.J.S. is supported by National Science Foundation
(NSF) grants 2308181, 2407566, and 2432036. The operation of the Konkoly Schmidt telescope is supported by the GINOP 2.3.2-15-2016-00003 grant, funded by the European Union. The supernova research group at Konkoly Observatory is supported by NKFIH-OTKA grant K142534, and GINOP grant 2.3.2-15-2016-00033.

M.M. and the METAL group at UVA acknowledge support in part from ADAP program grant  80NSSC22K0486, from NSF grant AST-2206657, and from the NSF under Cooperative Agreement 2421782 and the Simons Foundation grant MPS-AI-00010515 awarded to the NSF-Simons AI Institute for Cosmic Origins -- CosmicAI, https://www.cosmicai.org/. W.J.-G.\ is supported by NASA through Hubble Fellowship grant HSTHF2-51558.001-A awarded by the Space Telescope Science Institute, which is operated for NASA by the Association of Universities for Research in Astronomy, Inc., under contract NAS5-26555. S.W.J. gratefully acknowledges support from a Guggenheim Fellowship.
G.P.L. acknowledges support from the Royal Society (grants DHF-R1-221175 and DHF-ERE-221005). N.F. acknowledges support from the NSF Graduate Research Fellowship Program under grant DGE-2137419. A.R.L. acknowledges the support of grant 2025/09544-0 from the São Paulo Research Foundation (FAPESP).
A.V.F. is supported by many generous private donations.

This project used data obtained with the Dark Energy Camera (DECam), which was constructed by the Dark Energy Survey (DES) collaboration. Funding for the DES Projects has been provided by the DOE and NSF (USA), MISE (Spain), STFC (UK), HEFCE (UK), NCSA (UIUC), KICP (U. Chicago), CCAPP (Ohio State), MIFPA (Texas A\&M), CNPQ, FAPERJ, FINEP (Brazil), MINECO (Spain), DFG (Germany) and the Collaborating Institutions in the Dark Energy Survey, which are Argonne Lab, UC Santa Cruz, University of Cambridge, CIEMAT-Madrid, University of Chicago, University College London, DES-Brazil Consortium, University of Edinburgh, ETH Zürich, Fermilab, University of Illinois, ICE (IEEC-CSIC), IFAE Barcelona, Lawrence Berkeley Lab, LMU München and the associated Excellence Cluster Universe, University of Michigan, NSF NOIRLab, University of Nottingham, Ohio State University, OzDES Membership Consortium, University of Pennsylvania, University of Portsmouth, SLAC National Lab, Stanford University, University of Sussex, and Texas A\&M University. 
%G.P.L. acknowledges support from the Royal Society (grants DHF-R1-221175 and DHF-ERE-221005). N.F. acknowledges support from the NSF Graduate Research Fellowship Program under grant DGE-2137419. A.R.L. acknowledges the support of grant 2025/09544-0 from the São Paulo Research Foundation (FAPESP).

Based in part on observations at Cerro Tololo Inter-American Observatory, NSF NOIRLab (Prop. ID 2025B-589422; PI D. Sand), which is managed by the Association of Universities for Research in Astronomy (AURA) under a cooperative agreement with the U.S. NSF.
Some of the observations reported here were obtained at the MMT Observatory, a joint facility of the Smithsonian Institution and the University of Arizona.
This work makes use of observations from the Las Cumbres Observatory network. 
The LCO team is supported by NSF grants AST-2308113 and AST-1911151.

Based in part on observations obtained at the international Gemini Observatory, a program of NSF NOIRLab, which is managed by the Association of Universities for Research in Astronomy (AURA) under a cooperative agreement with the U.S. NSF on behalf of the Gemini Observatory partnership: the U.S. NSF (United States), National Research Council (Canada), Agencia Nacional de Investigaci\'{o}n y Desarrollo (Chile), Ministerio de Ciencia, Tecnolog\'{i}a e Innovaci\'{o}n (Argentina), Minist\'{e}rio da Ci\^{e}ncia, Tecnologia, Inova\c{c}\~{o}es e Comunica\c{c}\~{o}es (Brazil), and Korea Astronomy and Space Science Institute (Republic of Korea).

Based in part on observations obtained with the Hobby-Eberly Telescope (HET), which is a joint project of the University of Texas at Austin, the Pennsylvania State University, Ludwig-Maximillians-Universitaet Muenchen, and Georg-August Universitaet Goettingen. The HET is named in honor of its principal benefactors, William P. Hobby and Robert E. Eberly. The Low Resolution
Spectrograph 2 (LRS2) was developed and funded by
the University of Texas at Austin McDonald Observatory and Department of Astronomy and by Pennsylvania State University. 

The Pan-STARRS1 Surveys (PS1) and the PS1 public science archive have been made possible through contributions by the Institute for Astronomy, the University of Hawaii, the Pan-STARRS Project Office, the Max-Planck Society and its participating institutes, the Max Planck Institute for Astronomy, Heidelberg and the Max Planck Institute for Extraterrestrial Physics, Garching, The Johns Hopkins University, Durham University, the University of Edinburgh, the Queen's University Belfast, the Harvard-Smithsonian Center for Astrophysics, the Las Cumbres Observatory Global Telescope Network Incorporated, the National Central University of Taiwan, the Space Telescope Science Institute, the National Aeronautics and Space Administration under grant  NNX08AR22G issued through the Planetary Science Division of the NASA Science Mission Directorate,  the National Science Foundation grant AST–1238877, the University of Maryland, Eotvos Lorand University (ELTE), the Los Alamos National Laboratory, and the Gordon and Betty Moore Foundation.

\end{acknowledgments}

%%This section gives authors the space to recognize author contributions. The text inside this environment is NOT counted towards the total word quanta. At a minimum, manuscripts are expected to include this text:

%% But authors are expected to provide more specific details, e.g. 
%%
%%SC was responsible for writing and submitting the manuscript.
%%WWM came up with the initial research concept and edited the manuscript.
%%OTS obtained the funding and edited the manuscript.
%%EBF provided the formal analysis and validation. He also edited the manuscript.
%%GEH Supervised the undergraduates, wrote the software and administers the project github and Zenodo repositories.
%%
%% Authors can use the Contributor Role Taxonomy (CRediT) at
%% for ideas on how write a good statement tailored to their needs.

\facilities{CTIO:DECam, MMT:Bluechannel, MMT:Binospec, HET:LRS2, Gemini-S:GMOS, LCO, Konkoly}

\software{astropy \citep{2013A&A...558A..33A,2018AJ....156..123A,2022ApJ...935..167A},  
          Source Extractor \citep{1996A&AS..117..393B},
          SFFT \citep[][]{Hu_2022}, numpy \citep[][]{numpy}, pandas \citep[][]{pandas}, scipy \citep[][]{scipy}, IRAF \citep[][]{IRAF}, lcogtsnpipe \citep[][]{Valenti_2016} }


%\appendix

\bibliography{sample701}{}
\bibliographystyle{aasjournalv7}



\end{document}